\documentclass[manuscript, screen, nonacm]{acmart}

\usepackage{hyperref}
\usepackage{enumitem}
\usepackage{csquotes}
\usepackage{graphicx}
\usepackage{amsmath}
\usepackage{natbib}
\usepackage{xspace}
\usepackage{array,booktabs}
\usepackage[nolist]{acronym}
\usepackage{subcaption} 
\usepackage{adjustbox} 
\usepackage{calc}
\usepackage{caption}
\usepackage{multirow}
\usepackage{tabularx}

\usepackage{tikz}
\usetikzlibrary{
  arrows.meta,
  positioning,
  shapes.geometric,
  fit,
  backgrounds
}

\setlist[enumerate,1]{label=(\arabic*)}
\setlist[enumerate,2]{label=(\theenumi.\arabic*)}
\setlist[enumerate,3]{label=(\theenumii.\arabic*)}

\setlist[enumerate,1]{label=(\arabic*), ref=\arabic*}
\setlist[enumerate,2]{label=(\theenumi.\arabic*), ref=\theenumi.\arabic*}
\setlist[enumerate,3]{label=(\theenumii.\arabic*), ref=\theenumii.\arabic*}

\newcounter{qcounter}
\newcommand{\mychoice}[1]{{$\circ$}~#1 \, } 

\begin{document}

\begin{acronym}
    \acro{AI}[\textsc{AI}]{artificial intelligence}
    \acro{CNN}{convolutional neural network}
    \acro{DL}{deep learning}
    \acro{DNN}{deep neural network}
    \acro{ML}{machine learning}
    \acro{SVM}{support-vector machine}
    \acro{XAI}{explainable AI}
\end{acronym}

\title[Are explainable AI (XAI) evaluation strategies aligned?]{Are explainable AI (XAI) evaluation strategies aligned? Comparing subjective, objective, and mathematical evaluation measures using saliency maps}

\author{Felix Kares}
\authornote{These authors contributed equally to this research and share the first authorship.}
\email{felix.kares@uni-saarland.de}
\orcid{0009-0004-2046-2752}
\affiliation{%
  \institution{Saarland University}
  \city{Saarbr\"{u}cken}
  \country{Germany}
}

\author{Timo Speith}
\authornotemark[1]
\email{timo.speith@uni-bayreuth.de}
\orcid{0000-0002-6675-154X}
\affiliation{%
  \institution{University of Bayreuth}
  \city{Bayreuth}
  \country{Germany}
}

\author{Hanwei Zhang}
\authornotemark[1]
\email{zhang@depend.uni-saarland.de}
\orcid{0000-0002-9690-6952}
\affiliation{%
  \institution{Saarland University}
  \city{Saarbr\"{u}cken}
  \country{Germany}
}

\author{Markus Langer}
\email{markus.langer@psychologie.uni-freiburg.de}
\orcid{0000-0002-8165-1803}
\affiliation{%
  \institution{University of Freiburg}
  \city{Freiburg}
  \country{Germany}
}


\begin{acronym}
    \acro{AI}[\textsc{AI}]{artificial intelligence}
    \acro{CNN}{convolutional neural network}
    \acro{DL}{deep learning}
    \acro{DNN}{deep neural network}
    \acro{ML}{machine learning}
    \acro{SVM}{support-vector machine}
    \acro{XAI}{explainable AI}
\end{acronym}

\begin{abstract}
The evaluation of explainable AI (XAI) approaches often relies on three families of methods: subjective measures (e.g., questionnaires on trust or satisfaction), objective measures (e.g., task performance metrics), and mathematical metrics (e.g., for faithfulness). Yet, it remains unclear how these families align or diverge in practice.
In a{preregistered} between-subjects study (\textit{N}=166), we use three established saliency map techniques (LIME, Grad-CAM, Guided Backpropagation) as a testbed to examine this issue.
We find that each family of methods leads to different conclusions: participants reported no differences in trust or satisfaction, Grad-CAM improved user performance, while mathematical metrics favored Guided Backpropagation. At the same time, mathematical metrics were only partially related to user performance, and these relationships were sometimes counterintuitive. Our findings highlight the methodological importance of comparing subjective, objective, and mathematical approaches when evaluating XAI, illustrating both tensions and aspects that are aligned. We discuss implications for XAI evaluation frameworks.
\end{abstract}



\maketitle

\newcommand{\relu}{\operatorname{relu}}
\newcommand{\gap}{\operatorname{GAP}}
\newcommand{\up}{\operatorname{up}}

\newcommand{\cam}{\textrm{CAM}}
\newcommand{\gcam}{\textrm{Grad-CAM}}
\newcommand{\scam}{\textrm{Score-CAM}}

\newcommand{\Fdef}{Mask\xspace}
\newcommand{\Fref}{Diff\xspace}
\newcommand{\MIOFref}{IODiff\xspace}
\newcommand{\MIODref}{IOMask\xspace}

\newcommand{\AG}{\operatorname{AG}}
\newcommand{\AGf}{Average Gain\xspace}
\newcommand{\Agf}{Average gain\xspace}
\newcommand{\agf}{average gain\xspace}

\newcommand{\AC}{\operatorname{AC}}
\newcommand{\ACf}{Average Contract\xspace}

\newcommand{\AD}{\operatorname{AD}}
\newcommand{\I}{\operatorname{I}}
\newcommand{\D}{\operatorname{D}}
\newcommand{\AI}{\operatorname{AI}}
\newcommand{\OM}{\operatorname{OM}}
\newcommand{\LE}{\operatorname{LE}}
\newcommand{\Fo}{\operatorname{F1}}
\newcommand{\prc}{\operatorname{precision}}
\newcommand{\rec}{\operatorname{recall}}
\newcommand{\BA}{\operatorname{BoxAcc}}
\newcommand{\spg}{\operatorname{SP}}
\newcommand{\epg}{\operatorname{EP}}
\newcommand{\SM}{\operatorname{SM}}
\newcommand{\iou}{\operatorname{IoU}}

\newcommand{\head}[1]{{\smallskip\noindent\textbf{#1}}}
\newcommand{\alert}[1]{{\color{red}{#1}}}
\newcommand{\sm}{\scriptsize}
\newcommand{\eq}[1]{(\ref{eq:#1})}

\newcommand{\Th}[1]{\textsc{#1}}
\newcommand{\mr}[2]{\multirow{#1}{*}{#2}}
\newcommand{\mc}[2]{\multicolumn{#1}{c}{#2}}
\newcommand{\tb}[1]{\textbf{#1}}
\newcommand{\ch}{\checkmark}

\newcommand{\red}[1]{{\textcolor{red}{#1}}}
\newcommand{\blue}[1]{{\textcolor{blue}{#1}}}
\newcommand{\green}[1]{{\textcolor{green}{#1}}}

\newcommand{\citeme}[1]{\red{[XX]}}
\newcommand{\refme}[1]{\red{(XX)}}

\newcommand{\fig}[2][1]{\includegraphics[width=#1\linewidth]{fig/#2}}
\newcommand{\figh}[2][1]{\includegraphics[height=#1\linewidth]{fig/#2}}
\newcommand{\figa}[2][1]{\includegraphics[width=#1]{fig/#2}}
\newcommand{\figah}[2][1]{\includegraphics[height=#1]{fig/#2}}

\newcommand{\tran}{^\top}
\newcommand{\mtran}{^{-\top}}
\newcommand{\zcol}{\mathbf{0}}
\newcommand{\zrow}{\zcol\tran}

\newcommand{\ind}{\mathbbm{1}}
\newcommand{\expect}{\mathbb{E}}
\newcommand{\nat}{\mathbb{N}}
\newcommand{\zahl}{\mathbb{Z}}
\newcommand{\real}{\mathbb{R}}
\newcommand{\proj}{\mathbb{P}}
\newcommand{\prob}{\operatorname{P}}
\newcommand{\normal}{\mathcal{N}}

\newcommand{\mif}{\textrm{if}\ }
\newcommand{\other}{\textrm{otherwise}}
\newcommand{\minimize}{\textrm{minimize}\ }
\newcommand{\maximize}{\textrm{maximize}\ }
\newcommand{\st}{\textrm{subject\ to}\ }

\newcommand{\id}{\operatorname{id}}
\newcommand{\const}{\operatorname{const}}
\newcommand{\sgn}{\operatorname{sgn}}
\newcommand{\var}{\operatorname{Var}}
\newcommand{\mean}{\operatorname{mean}}
\newcommand{\trace}{\operatorname{tr}}
\newcommand{\diag}{\operatorname{diag}}
\newcommand{\vect}{\operatorname{vec}}
\newcommand{\cov}{\operatorname{cov}}
\newcommand{\sign}{\operatorname{sign}}
\newcommand{\prj}{\operatorname{proj}}

\newcommand{\softmax}{\operatorname{softmax}}
\newcommand{\argmax}{\operatorname{arg}\max}
\newcommand{\clip}{\operatorname{clip}}

\newcommand{\defn}{\mathrel{:=}}
\newcommand{\peq}{\mathrel{+\!=}}
\newcommand{\meq}{\mathrel{-\!=}}

\newcommand{\paren}[1]{\left({#1}\right)}
\newcommand{\mat}[1]{\left[{#1}\right]}
\newcommand{\floor}[1]{\left\lfloor{#1}\right\rfloor}
\newcommand{\ceil}[1]{\left\lceil{#1}\right\rceil}
\newcommand{\inner}[1]{\left\langle{#1}\right\rangle}
\newcommand{\norm}[1]{\left\|{#1}\right\|}
\newcommand{\abs}[1]{\left|{#1}\right|}
\newcommand{\frob}[1]{\norm{#1}_F}
\newcommand{\card}[1]{\left|{#1}\right|\xspace}

\newcommand{\diff}{\mathrm{d}}
\newcommand{\der}[3][]{\frac{\diff^{#1}#2}{\diff#3^{#1}}}
\newcommand{\ider}[3][]{\diff^{#1}#2/\diff#3^{#1}}
\newcommand{\pder}[3][]{\frac{\partial^{#1}{#2}}{\partial{{#3}^{#1}}}}
\newcommand{\ipder}[3][]{\partial^{#1}{#2}/\partial{#3^{#1}}}
\newcommand{\dder}[3]{\frac{\partial^2{#1}}{\partial{#2}\partial{#3}}}

\newcommand{\wb}[1]{\overline{#1}}
\newcommand{\wt}[1]{\widetilde{#1}}

\def\xssp{\hspace{1pt}}
\def\ssp{\hspace{3pt}}
\def\msp{\hspace{5pt}}
\def\lsp{\hspace{12pt}}

\newcommand{\cA}{\mathcal{A}}
\newcommand{\cB}{\mathcal{B}}
\newcommand{\cC}{\mathcal{C}}
\newcommand{\cD}{\mathcal{D}}
\newcommand{\cE}{\mathcal{E}}
\newcommand{\cF}{\mathcal{F}}
\newcommand{\cG}{\mathcal{G}}
\newcommand{\cH}{\mathcal{H}}
\newcommand{\cI}{\mathcal{I}}
\newcommand{\cJ}{\mathcal{J}}
\newcommand{\cK}{\mathcal{K}}
\newcommand{\cL}{\mathcal{L}}
\newcommand{\cM}{\mathcal{M}}
\newcommand{\cN}{\mathcal{N}}
\newcommand{\cO}{\mathcal{O}}
\newcommand{\cP}{\mathcal{P}}
\newcommand{\cQ}{\mathcal{Q}}
\newcommand{\cR}{\mathcal{R}}
\newcommand{\cS}{\mathcal{S}}
\newcommand{\cT}{\mathcal{T}}
\newcommand{\cU}{\mathcal{U}}
\newcommand{\cV}{\mathcal{V}}
\newcommand{\cW}{\mathcal{W}}
\newcommand{\cX}{\mathcal{X}}
\newcommand{\cY}{\mathcal{Y}}
\newcommand{\cZ}{\mathcal{Z}}

\newcommand{\vA}{\mathbf{A}}
\newcommand{\vB}{\mathbf{B}}
\newcommand{\vC}{\mathbf{C}}
\newcommand{\vD}{\mathbf{D}}
\newcommand{\vE}{\mathbf{E}}
\newcommand{\vF}{\mathbf{F}}
\newcommand{\vG}{\mathbf{G}}
\newcommand{\vH}{\mathbf{H}}
\newcommand{\vI}{\mathbf{I}}
\newcommand{\vJ}{\mathbf{J}}
\newcommand{\vK}{\mathbf{K}}
\newcommand{\vL}{\mathbf{L}}
\newcommand{\vM}{\mathbf{M}}
\newcommand{\vN}{\mathbf{N}}
\newcommand{\vO}{\mathbf{O}}
\newcommand{\vP}{\mathbf{P}}
\newcommand{\vQ}{\mathbf{Q}}
\newcommand{\vR}{\mathbf{R}}
\newcommand{\vS}{\mathbf{S}}
\newcommand{\vT}{\mathbf{T}}
\newcommand{\vU}{\mathbf{U}}
\newcommand{\vV}{\mathbf{V}}
\newcommand{\vW}{\mathbf{W}}
\newcommand{\vX}{\mathbf{X}}
\newcommand{\vY}{\mathbf{Y}}
\newcommand{\vZ}{\mathbf{Z}}

\newcommand{\va}{\mathbf{a}}
\newcommand{\vb}{\mathbf{b}}
\newcommand{\vc}{\mathbf{c}}
\newcommand{\vd}{\mathbf{d}}
\newcommand{\ve}{\mathbf{e}}
\newcommand{\vf}{\mathbf{f}}
\newcommand{\vg}{\mathbf{g}}
\newcommand{\vh}{\mathbf{h}}
\newcommand{\vi}{\mathbf{i}}
\newcommand{\vj}{\mathbf{j}}
\newcommand{\vk}{\mathbf{k}}
\newcommand{\vl}{\mathbf{l}}
\newcommand{\vm}{\mathbf{m}}
\newcommand{\vn}{\mathbf{n}}
\newcommand{\vo}{\mathbf{o}}
\newcommand{\vp}{\mathbf{p}}
\newcommand{\vq}{\mathbf{q}}
\newcommand{\vr}{\mathbf{r}}
\newcommand{\Vs}{\mathbf{s}}
\newcommand{\vt}{\mathbf{t}}
\newcommand{\vu}{\mathbf{u}}
\newcommand{\vw}{\mathbf{w}}
\newcommand{\vx}{\mathbf{x}}
\newcommand{\vy}{\mathbf{y}}
\newcommand{\vz}{\mathbf{z}}

\newcommand{\vone}{\mathbf{1}}
\newcommand{\vzero}{\mathbf{0}}

\newcommand{\valpha}{{\boldsymbol{\alpha}}}
\newcommand{\vbeta}{{\boldsymbol{\beta}}}
\newcommand{\vgamma}{{\boldsymbol{\gamma}}}
\newcommand{\vdelta}{{\boldsymbol{\delta}}}
\newcommand{\vepsilon}{{\boldsymbol{\epsilon}}}
\newcommand{\vzeta}{{\boldsymbol{\zeta}}}
\newcommand{\veta}{{\boldsymbol{\eta}}}
\newcommand{\vtheta}{{\boldsymbol{\theta}}}
\newcommand{\viota}{{\boldsymbol{\iota}}}
\newcommand{\vkappa}{{\boldsymbol{\kappa}}}
\newcommand{\vlambda}{{\boldsymbol{\lambda}}}
\newcommand{\vmu}{{\boldsymbol{\mu}}}
\newcommand{\vnu}{{\boldsymbol{\nu}}}
\newcommand{\vxi}{{\boldsymbol{\xi}}}
\newcommand{\vomikron}{{\boldsymbol{\omikron}}}
\newcommand{\vpi}{{\boldsymbol{\pi}}}
\newcommand{\vrho}{{\boldsymbol{\rho}}}
\newcommand{\vsigma}{{\boldsymbol{\sigma}}}
\newcommand{\vtau}{{\boldsymbol{\tau}}}
\newcommand{\vupsilon}{{\boldsymbol{\upsilon}}}
\newcommand{\vphi}{{\boldsymbol{\phi}}}
\newcommand{\vchi}{{\boldsymbol{\chi}}}
\newcommand{\vpsi}{{\boldsymbol{\psi}}}
\newcommand{\vomega}{{\boldsymbol{\omega}}}

\newcommand{\rLambda}{\mathrm{\Lambda}}
\newcommand{\rSigma}{\mathrm{\Sigma}}

\newcommand{\vLambda}{\bm{\rLambda}}
\newcommand{\vSigma}{\bm{\rSigma}}

\makeatletter
\newcommand*\bdot{\mathpalette\bdot@{.7}}
\newcommand*\bdot@[2]{\mathbin{\vcenter{\hbox{\scalebox{#2}{$\m@th#1\bullet$}}}}}
\makeatother

\makeatletter
\DeclareRobustCommand\onedot{\futurelet\@let@token\@onedot}
\def\@onedot{\ifx\@let@token.\else.\null\fi\xspace}

\def\eg{\emph{e.g}\onedot} \def\Eg{\emph{E.g}\onedot}
\def\ie{\emph{i.e}\onedot} \def\Ie{\emph{I.e}\onedot}
\def\cf{\emph{cf}\onedot} \def\Cf{\emph{Cf}\onedot}
\def\etc{\emph{etc}\onedot} \def\vs{\emph{vs}\onedot}
\def\wrt{w.r.t\onedot} \def\dof{d.o.f\onedot} \def\aka{a.k.a\onedot}
\def\etal{\emph{et al}\onedot}
\makeatother

\section{Introduction}

Enhancing user desiderata (e.g.,~satisfaction or understanding) is a key goal of \ac{XAI} approaches like saliency maps \citep{Langer2021What,Chazette2021Exploring}.  
Achieving this goal requires consistent evaluation and comparison of different approaches.
One popular family of evaluation methods is user studies \citep{Lapuschkin2016Analyzing,Ribeiro2016Why,Alqaraawi2020Evaluating}, where approaches are evaluated either subjectively through questionnaires
\citep{Hoffman2018Metrics}, or objectively, using user abilities or task performance \citep{Speith2024Conceptualizing,Mueller2025Explainable}. 
A more formalized family of evaluation methods focuses on certain properties (e.g.,~fidelity) that can be assessed via mathematical metrics \citep{Bylinskii2019What,Gomez2022Metrics,Kuemmerer2018Saliency,Speith2026Review}. 

In evaluating \ac{XAI} approaches, some researchers view mathematical metrics as a complement to user studies \citep{DoshiVelez2017Towards}, whereas others hope that these metrics could partially even replace user studies
\citep{Arya2019One,carvalho2019machine,Batic2024Toward}.  
This hope is partly motivated by the limitations of user studies.
For example, their results can be affected by individual biases, such as change blindness or a preference for usability over performance \citep{bertrand2022cognitive}. Moreover, user studies are resource-intensive and
therefore often difficult to scale.
{In contrast, the computation of mathematical metrics is easy to scale, typically standardized, and reproducible.} 
They can be used across large datasets and diverse models, allowing for widespread benchmarking of XAI approaches. 
However, mathematical metrics also have important limitations. 
For example, different metrics intended to measure the same property can diverge \citep{zhang2024saliency}, leading to contradictory evaluations.
Furthermore, user studies are indispensable for capturing human perspectives on desiderata such as trust and satisfaction, and for evaluating whether XAI approaches really lead to better understanding and better abilities in tasks where AI is used.

Although much research has evaluated XAI approaches like saliency maps with user studies \emph{or} mathematical metrics, articles often focus on only one explanation approach \citep{Mueller2025Explainable} or one
evaluation method \citep{zhang2024saliency}. 
This leaves open the more general question of how different families of evaluation methods relate to one another. 
Without explicitly comparing these perspectives within the same experimental context, we risk overlooking systematic divergences that matter for the evaluation of \ac{XAI} approaches more broadly. 
Moreover, the extent to which mathematical metrics can complement or even replace user studies has been explored only minimally, and usually for isolated cases. 

We contribute to this methodological debate by using three popular saliency maps that have different visual properties as a testbed for comparing evaluation families in practice.
{Image classification is a canonical task in XAI research~\citep{zhang2025saliency}, as it enables well-defined user tasks and objective performance measures while remaining broadly representative of real-world
applications.
In particular, saliency maps are among the most widely used XAI approaches, and they are frequently evaluated using both user studies and mathematical metrics, making them a suitable test case for comparing evaluation
paradigms~\citep{Speith2022Review}.} 

Specifically, we conducted a {preregistered} between-subjects online experiment where participants estimated an image classifier's accuracy based on saliency maps generated with Gradient-Weighted Class Activation
Mapping (Grad-CAM;  \citep{Selvaraju2017Grad-CAM}), Guided Backpropagation (GBP;  \citep{Springenberg2015Striving}), or Local Interpretable Model-Agnostic Explanations (LIME; \citep{Ribeiro2016Why}). 
Furthermore, we systematically evaluated saliency maps produced by the three approaches using a set of mathematical metrics. Beyond the standard practice in XAI methodology papers of comparing dataset-level mean metric values, we also checked the statistical significance ($p$-value) of these metrics, providing a complementary and more rigorous perspective on their effectiveness. {To ensure the validity of our study, we employed rigorous comprehension checks and test trials to ensure every participant had a strong understanding of saliency maps and the task at hand.}
We see three main contributions of this paper: 

\begin{enumerate}[label={(\arabic*)}]
\item We use saliency maps as a testbed to systematically compare evaluation methods for XAI approaches across families, assessing user reactions (trust, satisfaction), user abilities (task performance), and mathematical metrics (for fidelity, complexity, robustness, and localization) within the same experimental setting.
\item We show that user studies and mathematical metrics do not align: the approach rated most favorable on mathematical metrics (viz., GBP) was not the one that best supported user performance (viz., Grad-CAM). This illustrates the complementarity, but also the tension, between families of evaluation methods.
\item We show that while some mathematical metrics may correlate with user performance, these relationships can be counterintuitive and difficult to interpret. Together with (2), this challenges the assumption that mathematical metrics can serve as straightforward substitutes for user studies.
\end{enumerate}

\section{Background and Related Work}

\subsection{{{AI}} Image Classification and Saliency Maps}

Image classification is a task where AI models categorize images into predefined classes. 
It is widely used in fields such as healthcare (e.g.,~for diagnosing medical images; \citep{Pinto-Coelho2023How, Arun2021Assessing}), security (e.g.,~for facial recognition; \citep{Girmay2021AI,
Rathi2023AI}), autonomous vehicles (e.g.,~for object detection; \citep{Ma2020Artificial}), and retail (e.g.,~for visual search engines; \citep{Oosthuizen2021Artificial}). 
Research on image classification includes improving model accuracy \citep{rawat2017deep}, augmenting adversarial robustness \citep{dong2020benchmarking}, and reducing bias \citep{mehrabi2021survey}.
{Convolutional neural networks (CNNs) have achieved remarkable success in image classification~\citep{Krizhevsky2012ImageNet} and remain widely used in various computer vision tasks
\citep{younesi2024comprehensive,reza2026comprehensive}.}

With the use of AI in sensitive areas such as healthcare, trustworthiness has become increasingly important \citep{Li2022Interpretable}. 
\Ac{XAI} aims to contribute to assessing the trustworthiness of AI-based systems, for example, by finding out whether different types of explanations can enable users to better understand classifiers or better assess which
outputs are correct and which are incorrect \citep{Lim2009Why}. 
Likewise, XAI also involves empowering users to make better decisions based on AI outputs \citep{Baum2022Responsibility} and can contribute to more favorable reactions towards classifiers as people may find them more
helpful \citep{pesecan2023increasing}.

Saliency maps are a popular explanation for AI image classification \citep{Speith2022Review}, highlighting image regions most influential in the AI model's decision-making process, to visually represent what the AI
\enquote{sees} as important.
Despite rapid changes in dominant AI paradigms, saliency maps remain foundational to XAI. The principles and insights established by saliency-based analysis continue to directly shape and support the development of XAI
methods for emerging architectures, novel tasks, and diverse input formats~\citep{mumuni2025explainable,dang2024explainable}.
In general, saliency maps aim to make the decision process of AI systems more transparent, allowing users to better understand the model and make informed decisions when it should be trusted \citep{Ribeiro2016Why}. 
At the same time, saliency maps can also help developers to identify and fix mistakes in AI models \citep{Folke2021Explainable}, thus providing a tool for laypersons and experts alike. 
In the context of this paper, we will highlight three popular methods of saliency mapping: Grad-CAM \citep{Selvaraju2017Grad-CAM}, GBP \citep{Springenberg2015Striving}, and LIME \citep{Ribeiro2016Why}.

We chose these three methods for generating saliency maps because each of them works differently and has its own visual properties. 
Specifically, Grad-CAM visualizes the \emph{areas} in the image that contribute most to the predicted class by leveraging the gradients flowing into the last convolutional layer. 
Visually, Grad-CAM highlights importance using different colors, for example, by using warmer colors for high importance areas and colder colors for low importance ones. 
LIME perturbs the input data and observes changes in the output to identify which parts of the image are most important. This results in individual \emph{sections} being separated and highlighted. 
Finally, GBP modifies the standard backpropagation process, allowing only positive gradients to contribute to the visualization, resulting in more focused feature maps. 
GBP saliency maps highlight mostly the \emph{edges} of the most important features used for classification.

\subsection{Evaluating {{XAI}} Approaches}

When it comes to evaluating \ac{XAI} approaches, a distinction is commonly made between human-centered evaluations and functionally grounded evaluations (i.e.,~mathematical metrics; see \citep{DoshiVelez2017Towards,
Zhou2021Evaluating, Vilone2021Notions, Alangari2023Exploring, Speith2026Review}). 
Within the human-centered perspective, researchers often distinguish between subjective evaluations (e.g.,~trust or satisfaction ratings) and objective evaluations (e.g.,~task performance or accuracy improvements; see
\citep{Vilone2021Notions, Zhou2021Evaluating}). 
This distinction (subjective, objective, mathematical) is widely recognized \citep{Speith2026Review}, yet how these different families of evaluation align or diverge remains an open question \citep{Zhou2021Evaluating}. 

Several theoretical frameworks have pointed out that relying on only one family of evaluation risks producing misleading conclusions \citep{DoshiVelez2017Towards,Zhou2021Evaluating}. 
For instance, explanations may increase subjective trust without necessarily improving objective performance \citep{kaur2020interpreting}. 
At the same time, empirical evidence directly comparing evaluation families remains limited. 
While some studies have examined subjective and objective evaluations together (e.g.,~\citep{kaur2020interpreting}), these typically focus on a single XAI approach and do not incorporate mathematical metrics. 
Most studies evaluate a single XAI approach with one type of metric or restrict themselves to one family of evaluation \citep{Vilone2021Notions,Zhou2021Evaluating}. 
As a result, we still lack systematic empirical evidence on how subjective, objective, and mathematical evaluations converge or diverge when applied to the same \ac{XAI} approach. 
We contribute to filling this gap by examining different types of saliency maps, which allows us to directly compare evaluation strategies across families within the same experimental setting.
In the next sections, we will briefly introduce each of the three evaluation families followed by how they could be connected.

\subsubsection{User Studies with Subjective Measures}

When it comes to user studies that use subjective measures to evaluate different \ac{XAI} approaches, trust and explanation satisfaction are important user reactions
\citep{Hoffman2018Metrics,Brdnik2023Assessing,Hoffman2023Measures}. 
Trust can be defined as \enquote{the willingness of a party to be vulnerable to the actions of another party based on the expectation that the other will perform a particular action important to the trustor, irrespective
of the ability to monitor or control that other party} \citep[p.~712]{Mayer1995}. 
Explanation satisfaction combines \enquote{the degree to which users feel that they understand the AI system or process being explained to them} \citep[p.~3]{Hoffman2018Metrics} and their \enquote{feeling of satisfaction}
\citep[p.~4]{Hoffman2018Metrics} with the explanation.  

Findings on the impact of \ac{XAI} approaches on user reactions such as trust and satisfaction are mixed. 
A recent review by \citet{Rong2024Towards} showed that only about half of the studies comparing explanations to no explanations or placebo (randomly generated) explanations found a positive effect on trust. 
In the context of saliency maps, comparative studies have yielded insights but also left open questions. 
For instance, \citet{Khadivpour2022Responsibility} and \citet{Ihongbe2024Evaluating} report that participants tend to prefer Grad-CAM over LIME, perceiving it as more understandable or trusting it more. 
However, trust and understandability in these studies are mostly measured by using single-item measures (e.g.,~asking participants to rank their preference, see \citep{Khadivpour2022Responsibility}).
While single-item measures can be an efficient choice to indicate rough inclinations in terms of preferences, they provide limited insights into potentially multifaceted constructs such as trust
\citep{Allen2022Single,sarstedt2009more}. 
To our knowledge, there is also no investigation into user perceptions of GBP. 
Through our study, we want to address these gaps in research and provide a more rigorous investigation of different methods in terms of user-perceived satisfaction and trust.
Against this backdrop, we ask the following open research questions (RQs)\footnote{ 
In the preregistration, we formulated the hypotheses that GBP would be preferred over the other two approaches both in terms of user reactions and abilities. 
We have based this hypothesis on considerations relating to human perception, which, according to our initial research, prefers edges over surfaces. 
However, we realized that the prior empirical evidence is more diverse than we initially thought. 
This is why we now propose open RQs.}:

\begin{description}[leftmargin=!, labelwidth=\widthof{RQ1.1}]
\item[\textbf{RQ1.1}] Are there differences between the three approaches in perceived explanation satisfaction?
\item[\textbf{RQ1.2}] Are there differences between the three approaches in trust in the classifier and in the saliency map?
\end{description}

\subsubsection{User Studies with Objective Measures}

Increasing user understanding through explanations is a key goal of \ac{XAI} research \citep{Chazette2021Exploring,Speith2024Conceptualizing}. 
For example, explanations help users understand and exert control over systems that could otherwise cause harm \citep{Baum2022Responsibility}. 
Through increased understanding, humans can stay in control of high-stakes decision scenarios involving automation \citep{Abdul2018Trends}. 
This ensures that humans remain accountable for the outcomes systems might produce.  

Building on \citet{Speith2024Conceptualizing}, we believe that a promising way to grasp whether different explanations enhance user understanding is to examine their impact on certain user abilities. 
For saliency maps as explanations, important and often measured abilities related to how users interpret the results, also called the \emph{perceptibility} of saliency maps \citep{Boggust2023Saliency}, are how accurately
users can predict AI accuracy and AI predictions \citep{Mueller2025Explainable}. 
This is because increased understanding of a system should enable users to predict its output \citep{Alqaraawi2020Evaluating,Speith2024Conceptualizing}. 
A way to measure this is by asking users what percentage of images would be classified correctly by an image classifier for a specific class of images, or by presenting images and asking which of them would be classified correctly by the classifier. 

{Even though foundational studies provide theoretical arguments on the perceptual relevance of edges and surfaces (e.g.,~\citep{palmer2008extremal}), which could motivate the expectation that GBP would
outperform Grad-CAM or LIME, the} empirical evidence on which saliency maps enhance user understanding best is mixed and has methodological limitations.
For example, \citet{Colin2022What} compare six saliency methods across three classifiers and show that users' prediction performance improves in some settings but not others.
Similarly, \citet{Khadivpour2022Responsibility} compare three saliency maps (Grad-CAM, LIME, and SHAP) as well as three other visual XAI approaches and report broadly comparable performance across saliency maps; however,
they use a small sample size of 20 participants with only five predictions per condition, limiting the statistical power to detect meaningful differences.
Relatedly, \citet{Nguyen2021Effectiveness} find no significant differences among saliency maps, and report that such maps do not reliably improve performance across tasks, and may even perform worse than non-saliency
approaches in some settings.
In contrast, \citet{morrison2023eye} found that Grad-CAM outperformed LIME in a game-based setting. 
Finally, \citet{Kim2022HIVE} showed that Grad-CAM enabled above-chance-level predictions in a broad comparison against non-saliency explanation strategies. However, they did not statistically compare user performance
across different saliency map approaches.

While prior research has explored how different saliency map approaches influence user understanding, it remains unclear whether one approach consistently outperforms the others. 
The comparison groups have varied considerably, and while Grad-CAM was part of most of them, approaches such as LIME and especially GBP (which does not appear at all in the above-mentioned studies) have been under-researched.
Our study fills this gap by systematically comparing Grad-CAM, LIME, and GBP using a robust sample size and varied evaluation methods, offering new insights into GBP's effectiveness. 
Given the lack of a clearly superior approach in the literature, we pose the following open RQs\footnote{We also measured which features participants believed the classifier used based on the saliency map in each trial
(following \citep{Alqaraawi2020Evaluating}). Since naming more features does not necessarily indicate better understanding, we do not report these results in detail. Participants in the GBP group detected more features
than those in the LIME group, with no differences for the other comparisons. Results are available on request.}:

\begin{description}[leftmargin=!, labelwidth=\widthof{RQ2.1}]
\item[\textbf{RQ2.1}] Are there differences between the three approaches in participants' accuracy prediction relative to the classifier's true accuracy?
\item[\textbf{RQ2.2}] Are there differences between the three approaches in participants' error rates when predicting whether individual images would be classified correctly by the classifier? 
\end{description}

\subsubsection{Mathematical Metrics}

Besides evaluation through studies with humans, mathematical metrics play a key role in evaluating \ac{XAI} approaches \citep{Bylinskii2019What,li2021experimental,Speith2026Review}. 
In the field of \ac{XAI}, mathematical metrics for evaluating saliency maps are primarily used to evaluate properties such as \emph{fidelity}~\citep{Gomez2022Metrics,han2022explanation},
\emph{robustness}~\citep{Samek2017Evaluating,dasgupta2022framework}, \emph{complexity}~\citep{Bhatt2020Evaluating,nguyen2020quantitative}, and \emph{localization}~\citep{li2021experimental,zhang2016top}.

\emph{Fidelity}, also known as faithfulness or correctness, describes an explanation's ability to accurately reflect the decision-making process of a black-box model~\citep{carvalho2019machine}. 
Fidelity is a crucial concept for an explanation since it is meant to state whether an explanation reproduces the dynamics of the underlying model~\citep{Alvarez-Melis2018Towards}.
In our user study, participants were asked to predict the model's behavior based on saliency maps. We therefore expected fidelity to be particularly important: if a saliency map does not faithfully reflect the model's decision process, participants' predictions should also be less accurate.
\emph{Robustness}, or sensitivity, expresses the consistency of explanations under input perturbations. 
Since users may struggle to understand how a classifier works when presented with inconsistent explanations for slightly varying inputs~\citep{chen2022does} and trust it less, we expect robustness to be relevant for human interpretation of saliency maps.
\emph{Complexity} expresses the conciseness of explanations, meaning that only a few features are needed to explain a model's prediction. We expect that concise explanations may be easier to understand because they focus on a few key regions rather than emphasizing everything. Additionally, users might be more satisfied with concise explanations, as these are easier to parse.
Finally, for image explanations, \emph{localization} captures the extent to which saliency maps align with human-annotated bounding boxes indicating object locations in an image. We expect localization to be relevant to human evaluation because highlighting semantically meaningful, object-related regions may help users interpret the model's behavior. However, whether localization metrics are associated with human evaluations remains an empirical question.

With various mathematical metrics designed to assess specific properties, key questions arise regarding their effectiveness in reflecting saliency map performance. 
For instance, while crucial in computer vision, \emph{localization} has been criticized for its inability to differentiate between errors in the saliency map and the model~\citep{li2021experimental}. 
More generally, \citet{zhang2024saliency} and \citet{li2021experimental} have conducted extensive experiments across metrics, primarily focusing on benchmarking. 
Both teams observed that no single saliency method outperforms others across all metrics. 
While Grad-CAM performed fairly well across most metrics, metrics for the same property sometimes diverged when comparing different saliency maps. 
Against the background of these inconclusive findings, we thus ask:

\begin{description}[leftmargin=!, labelwidth=\widthof{RQ3}]
\item[\textbf{RQ3}] Are there differences between the three approaches in mathematical metrics that measure the properties fidelity, robustness, complexity, and localization?
\end{description}

Some studies have explored whether mathematical metrics can be associated (e.g.,~as proxies) with user studies. 
\citet{Nguyen2021Effectiveness} and \citet{Kim2022HIVE}, for instance, found that localization metrics do not indicate how well humans can judge the correctness of predictions with the help of different saliency maps like
Grad-CAM. 
Similarly, \citet{Colin2022What} found that fidelity metrics are unrelated to the \enquote{utility} of a saliency map (i.e.,~how well participants anticipate a classifier's prediction based on diverse saliency maps). 

We aim to expand these findings by broadening the scope of metrics and saliency map approaches. 
Specifically, we examine multiple metrics representing different properties (fidelity, complexity, robustness, and localization) across several widely used saliency maps. 
To our knowledge, this is also the first study to investigate LIME and GBP in this context.  
We thus ask\footnote{As subjective user metrics can be error-prone, we focus on the relation between mathematical metrics and objective measures of user understanding, i.e.,~participants' accuracy predictions. {However, the association of mathematical metrics with subjective measures could, in principle, also be interesting. As described above, what is measured in the different metrics might impact feelings of trust and satisfaction (e.g.,~people might be more satisfied with concise explanations).}}:
\begin{description}[leftmargin=!, labelwidth=\widthof{RQ4}]
\item[\textbf{RQ4}] Are the mathematical metrics that measure the properties fidelity, robustness, complexity, and localization associated with how well participants predict the accuracy of different classes?
\end{description}

\section{Methods}

\subsection{Design and Sample}

We conducted an a priori power analysis for a one-way ANOVA with three groups using \textit{G*Power} \citep{faul2007gpower} for a one-factor (saliency map: Grad-CAM vs. GBP vs. LIME) between-participants design. To detect a
small to medium effect size of $f = .175$, with a power of $1 - \beta = .80$ and $\alpha=.05$, a sample of 174 participants was needed. 

\definecolor{navy}{HTML}{17324D}
\definecolor{blue}{HTML}{3A86FF}
\definecolor{teal}{HTML}{2A9D8F}
\definecolor{purple}{HTML}{7A5AA6}
\definecolor{orange}{HTML}{F28E2B}
\definecolor{lightblue}{HTML}{EAF2FB}
\definecolor{lightteal}{HTML}{E8F6F3}
\definecolor{lightpurple}{HTML}{F1ECF8}
\definecolor{lightorange}{HTML}{FFF1E6}
\definecolor{lightgray}{HTML}{F5F7FA}
\definecolor{border}{HTML}{CBD5E1}
\definecolor{dark}{HTML}{263238}

\tikzset{
  font=\sffamily,
  >={Latex[length=2.8mm,width=2mm]},
  flow/.style={
    draw=navy,
    rounded corners=2mm,
    very thick,
    fill=white,
    text=dark,
    align=center,
    minimum height=12mm,
    text width=66mm,
    inner sep=3mm
  },
  phase/.style={
    flow,
    fill=lightblue
  },
  decision/.style={
    diamond,
    draw=orange,
    very thick,
    fill=lightorange,
    text=dark,
    align=center,
    aspect=2.1,
    inner sep=1.5mm,
    text width=32mm
  },
  method/.style={
    draw=purple,
    rounded corners=2mm,
    thick,
    fill=lightpurple,
    text=dark,
    align=center,
    minimum height=10mm,
    text width=34mm,
    inner sep=2.5mm
  },
  trial/.style={
    draw=teal,
    rounded corners=2mm,
    very thick,
    fill=lightteal,
    text=dark,
    align=center,
    minimum height=12mm,
    text width=66mm,
    inner sep=3mm
  },
  arrow/.style={
    ->,
    very thick,
    draw=navy
  },
  feedback/.style={
    ->,
    thick,
    draw=orange,
    dashed
  },
  groupbox/.style={
    draw=border,
    rounded corners=3mm,
    thick,
    fill=lightgray,
    inner sep=5mm
  },
  labelbox/.style={
    fill=navy,
    text=white,
    rounded corners=1.5mm,
    font=\bfseries\small,
    inner xsep=3mm,
    inner ysep=1.5mm
  }
}

\begin{figure}
\centering
\resizebox{\linewidth}{!}{%
\begin{tikzpicture}[
  node distance=8mm and 28mm
]


\node[phase] (recruit) {
  Recruitment of English-speaking\\
  ICT graduates via Prolific
};

\node[phase, below=of recruit] (assign) {
  Random assignment to one\\
  saliency-map condition
};

\node[
  method,
  text width=28mm,
  below=12mm of assign
] (gbp) {
  Guided\\Backpropagation
};

\node[
  method,
  text width=28mm,
  left=5mm of gbp
] (gradcam) {
  Grad-CAM
};

\node[
  method,
  text width=28mm,
  right=5mm of gbp
] (lime) {
  LIME
};

\node[
  phase,
  below=14mm of gbp
] (instruction) {
  General introduction to saliency maps\\
  + method-specific interpretation instructions
};

\node[
  decision,
  below=10mm of instruction
] (check) {
  Three comprehension\\
  questions passed?
};

\node[
  phase,
  below=24mm of check
] (practice) {
  Two example trials explaining\\
  both participant tasks
};

\node[
  phase,
  below=of practice
] (confidence) {
  Confidence rating before\\
  the experimental trials
};


\node[
  trial,
  right=36mm of recruit
] (stimulus) {
  Inspect three images and their\\
  corresponding saliency maps
};

\node[
  trial,
  below=of stimulus
] (accuracy) {
  Task 1: estimate the classifier's\\
  accuracy for the image class
};

\node[
  trial,
  below=of accuracy
] (classify) {
  Task 2: judge which of six additional\\
  images will be classified correctly
};

\node[
  decision,
  below=10mm of classify
] (trials) {
  Twelve experimental\\
  trials completed?
};

\node[
  phase,
  below=24mm of trials
] (post) {
  Post-study questionnaires
};

\node[
  phase,
  below=of post
] (demo) {
  Demographics and\\
  final data consent
};


\draw[arrow] (recruit) -- (assign);

\draw[arrow]
  (assign.south)
  -- ++(0,-4mm)
  -| (gradcam.north);

\draw[arrow]
  (assign.south)
  -- (gbp.north);

\draw[arrow]
  (assign.south)
  -- ++(0,-4mm)
  -| (lime.north);

\draw[arrow]
  (gradcam.south)
  -- ++(0,-4mm)
  -| (instruction.north);

\draw[arrow]
  (gbp.south)
  -- (instruction.north);

\draw[arrow]
  (lime.south)
  -- ++(0,-4mm)
  -| (instruction.north);

\draw[arrow] (instruction) -- (check);

\draw[feedback]
  (check.west)
  -- ++(-12mm,0)
  |- node[pos=.25, left, font=\small] {No}
  (instruction.west);

\draw[arrow]
  (check)
  -- node[pos=.25, right, font=\small] {Yes}
  (practice);

\draw[arrow] (practice) -- (confidence);


\draw[arrow]
  (confidence.east)
  -- ++(27.4mm,0)
  |- (stimulus.west);


\draw[arrow] (stimulus) -- (accuracy);
\draw[arrow] (accuracy) -- (classify);
\draw[arrow] (classify) -- (trials);

\draw[feedback]
  (trials.east)
  -- ++(12mm,0)
  |- node[pos=.10, left, font=\small] {No}
  (stimulus.east);

\draw[arrow]
  (trials)
  -- node[pos=.25,right, font=\small] {Yes}
  (post);

\draw[arrow] (post) -- (demo);


\begin{scope}[on background layer]

  \node[
    groupbox,
    fit=(recruit)(assign)(gradcam)(gbp)(lime)(instruction)(check),
    inner xsep=7mm,
    inner ysep=11mm
  ] (onboardingbox) {};

  \node[
    groupbox,
    fit=(practice)(confidence),
    inner xsep=7mm,
    inner ysep=9mm,
    yshift=2mm
  ] (practicebox) {};

  \node[
    groupbox,
    fit=(stimulus)(accuracy)(classify)(trials),
    inner xsep=7mm,
    inner ysep=11mm
  ] (experimentbox) {};

  \node[
    groupbox,
    fit=(post)(demo),
    inner xsep=7mm,
    inner ysep=9mm,
    yshift=2mm
  ] (postbox) {};

\end{scope}


\node[labelbox, anchor=north west]
  at ([xshift=2mm,yshift=-2mm]onboardingbox.north west)
  {Onboarding and assignment};

\node[labelbox, anchor=north west]
  at ([xshift=2mm,yshift=-2mm]practicebox.north west)
  {Practice};

\node[labelbox, anchor=north west]
  at ([xshift=2mm,yshift=-2mm]experimentbox.north west)
  {Experimental procedure};

\node[labelbox, anchor=north west]
  at ([xshift=2mm,yshift=-2mm]postbox.north west)
  {Post-study measures};

\end{tikzpicture}%
}
\caption{{Flowchart of the experimental procedure.}}
\label{fig:flowchart}
\end{figure}

We implemented our questionnaire using \textit{SoSci Survey} and recruited English-speaking graduates in information and communication technology (ICT) via \textit{Prolific}, expecting them to better understand saliency maps and be more familiar with the topic than the general population. ICT graduates may (in the future) also be the ones who assess the quality of AI models. 
The study was approved by the Ethical Review Board of the university of one of the first authors. 

Of the 175 participants who completed all questions, we excluded one who indicated their data should not be used, six who indicated that they did not have at least a bachelor's or equivalent degree, and two who took less than 8 minutes to complete the survey. The final sample consisted of 166 participants (28.6\% female and 1.2\% diverse; 4.8\% with a PhD or equivalent, 27.98\% with a master's degree or equivalent; $M_{\text{age}} = 31.19$, $SD_{\text{age}} = 7.83$).

\subsection{Procedure}

Participants were randomly assigned to one of three groups (Grad-CAM, GBP, or LIME). They first read a text explaining saliency maps (see Section B.3), followed by a group-specific text on interpreting AI classifications with the assigned saliency map approach. Participants then answered three comprehension questions, each with four answer options of which only one was correct. Incorrect answers prompted a return to the explanation, and participants who answered incorrectly twice were screened out.

\begin{figure*}[!htbp]
    \centering
    \begin{minipage}[b]{0.325\textwidth}
        \centering
        \includegraphics[width=\textwidth]{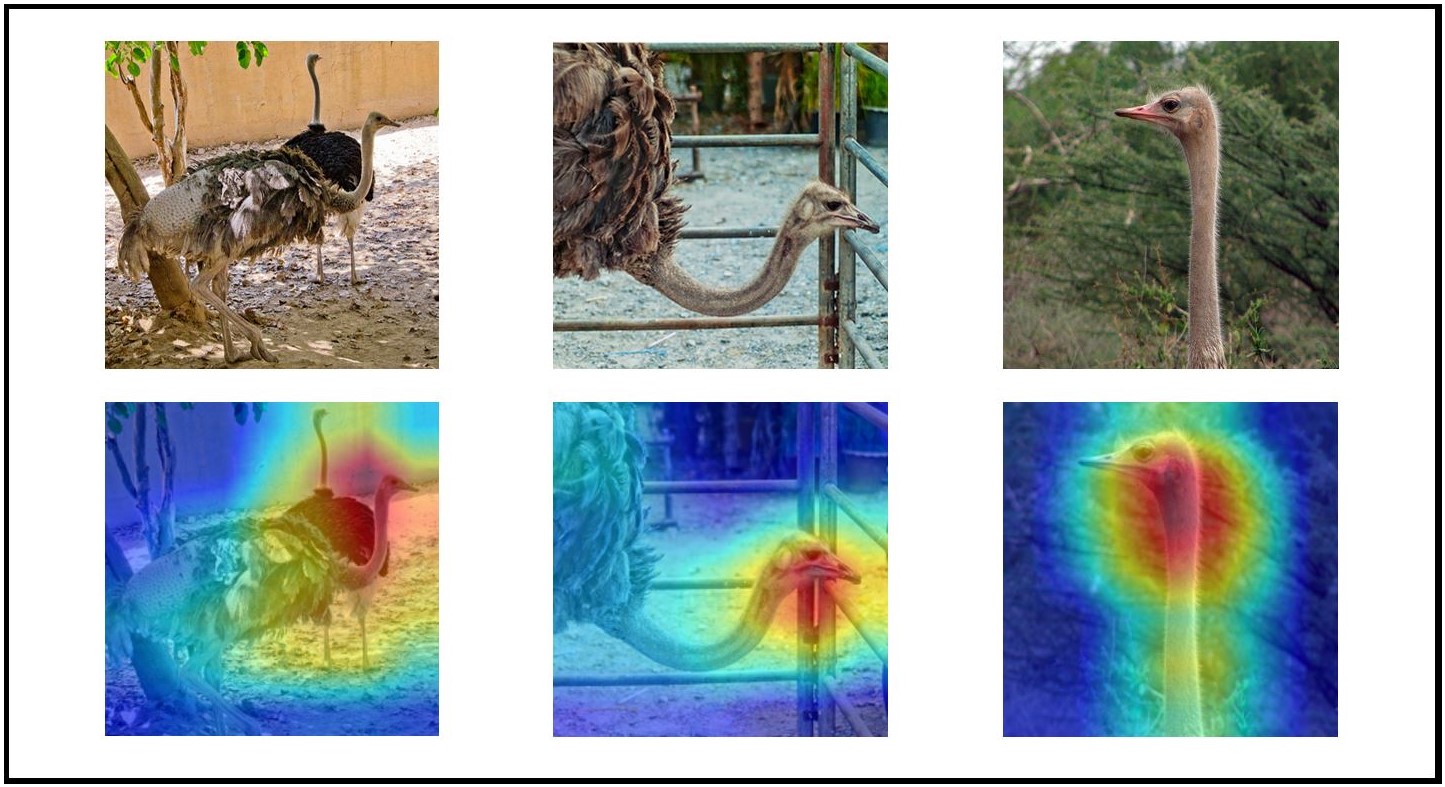}
        \subcaption{Grad-CAM}
        \label{fig:gradcam}
    \end{minipage}
    \hfill
    \begin{minipage}[b]{0.325\textwidth}
        \centering
        \includegraphics[width=\textwidth]{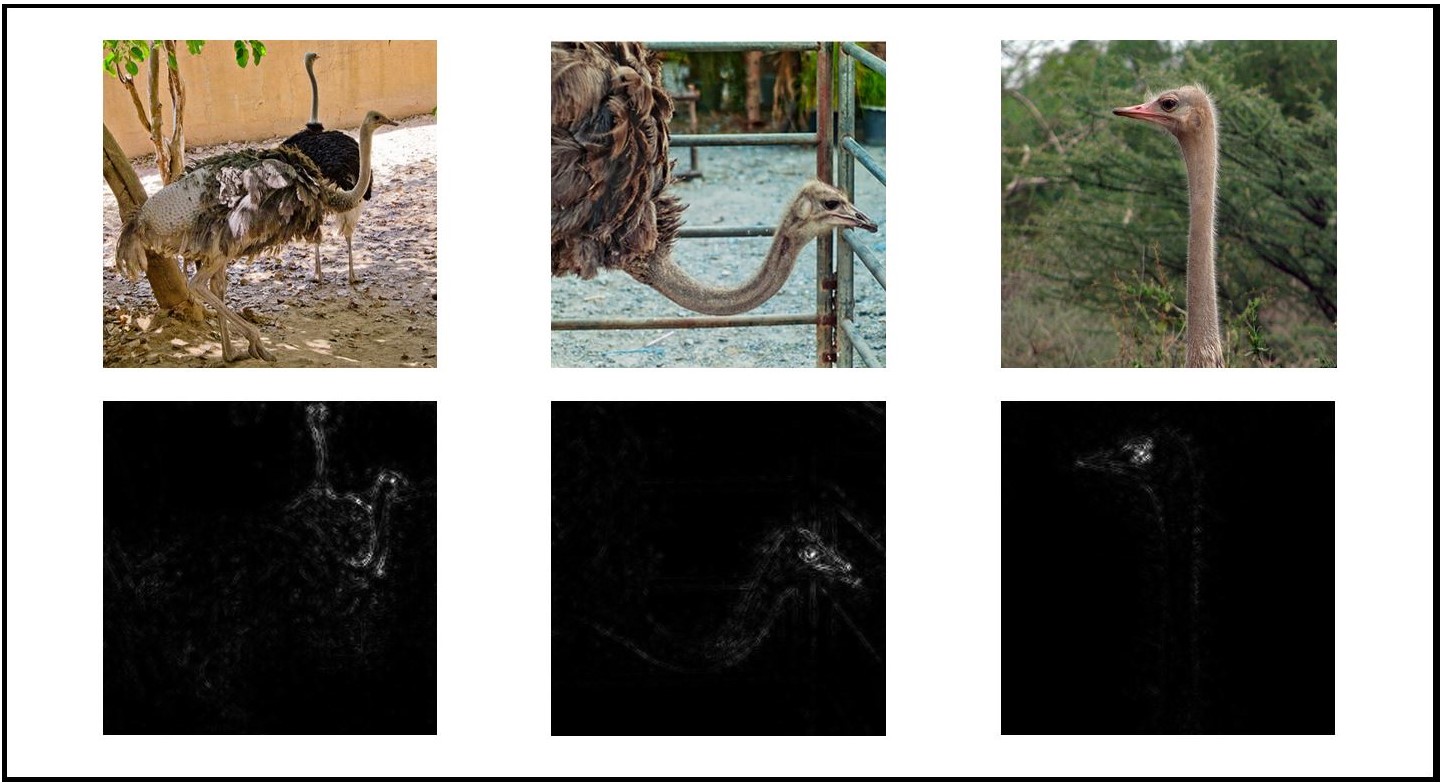}
        \subcaption{Guided Backpropagation}
        \label{fig:gbp}
    \end{minipage}
    \hfill
    \begin{minipage}[b]{0.325\textwidth}
        \includegraphics[width=\textwidth]{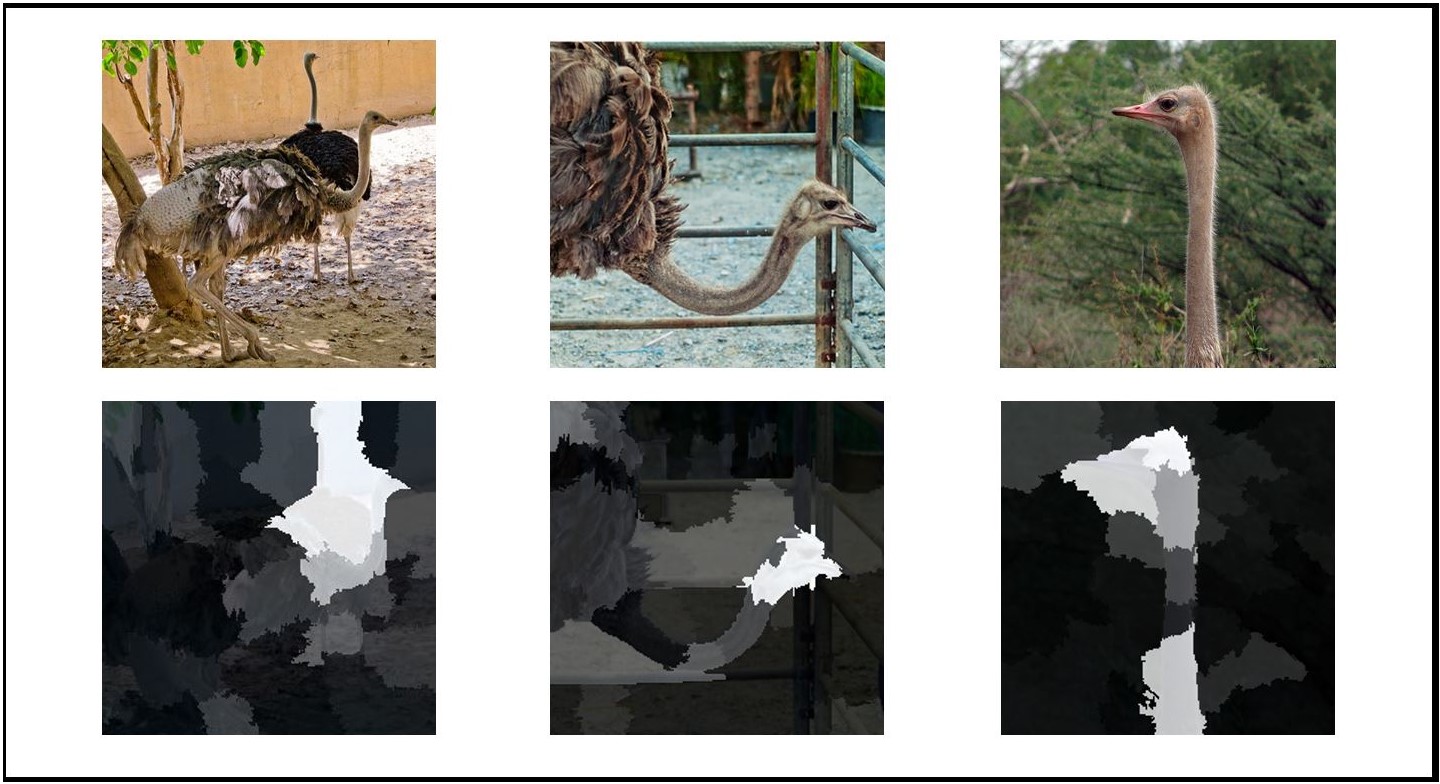}
    \subcaption{LIME}
    \label{fig:lime}
    \end{minipage}

    \caption{ \emph{Example trial images from the three groups.} Comparative visualization of saliency methods (Grad-CAM, GBP, LIME) using ostrich examples. Each section shows the original image (top) and corresponding saliency maps (bottom).}
    \label{fig:trialimages}
\end{figure*}

After correctly answering the comprehension questions, participants were shown two example trials followed by 12 experimental trials. In each trial, participants saw 12 images from the same class in two groups. The first group consisted of 3 images and 3 saliency maps generated from them (see \autoref{fig:trialimages}). The second group consisted of 6 further images without saliency maps (see Figure 7). 

The example trials explained the tasks to be performed:  (1) estimate the accuracy of the classifier for the respective class (using the first group of images), and (2) identify which of the six images from the second group of images would be correctly classified by the image classifier (see Section B.3.5).


After the practice trials, participants rated their confidence in evaluating a classifier using saliency maps, then completed the 12 experimental trials. Lastly, they answered questions regarding their explanation satisfaction, trust in the saliency maps, trust in and trustworthiness of the classifier, demographic questions, and data consent. The complete experimental procedure is visualized in \autoref{fig:flowchart}.

\subsection{Measures and Materials}

\subsubsection{Network, Dataset, and Saliency Maps}

We used the pretrained ResNet50~\citep{He2016Deep}, a classic CNN model, from the PyTorch model zoo.\footnote{\url{https://pytorch.org/vision/0.8/models.html}}  
Images were sampled from the ImageNet ILSVRC 2012 validation set, a large, diverse, well-known standardized benchmark dataset for image classification~\citep{Krizhevsky2012ImageNet} with $50,\!000$ images evenly
distributed across $1,\!000$ categories. 
We randomly chose 12 classes. Each trial included three images for the saliency maps and six images per class for participant classification, totaling {12~$\times$~9} images for the experimental trials and {2~$\times$~9} for practice (note that we do not count the saliency maps themselves here, as they are based on the first three images).

Grad-CAM~\citep{Selvaraju2017Grad-CAM} and GBP~\citep{Springenberg2015Striving} were implemented using \citeauthor{Gildenblat2021PyTorch}'s GitHub package \citep{Gildenblat2021PyTorch} while the implementation of LIME~\citep{Ribeiro2016Why} is from its official GitHub~\citep{Ribeiro2016Why}.

\subsubsection{Mathematical Metrics}\label{sec:mat:metrics}

We selected Insertion~\citep{petsiuk2018rise}, Deletion~\citep{petsiuk2018rise}, and Monotonicity~\citep{Arya2019One} as representative metrics for fidelity; Average Sensitivity (Avg.Sensitivity;~\citep{Yeh2019Fidelity}) for robustness; Entropy
Complexity (Ent.Complexity;~\citep{Bhatt2020Evaluating}) and Sparseness~\citep{Chalasani2020Concise} as two metrics for complexity; and Energy Pointing Game (EPG;~\citep{Wang2020Score-CAM}) for localization.
The implementation of Insertion and Deletion is from its official GitHub \citep{petsiuk2018rise}. The implementation of the other metrics is from the Quantus GitHub \citep{Hedstroem2023Quantus}.

Concerning fidelity, \emph{Insertion} and \emph{Deletion}~\citep{petsiuk2018rise} sequentially add or remove pixels in decreasing order of saliency and observe the effect on the prediction. Deletion measures the decrease
in the probability of a class when pixels are removed, with removal being equivalent to setting pixel values to zero (i.e.,~black). 
Conversely, Insertion measures the increase in probability when pixels are gradually reintroduced to a blurred version of the image (created by applying a Gaussian blur to the original image). 
Similar to Insertion, \emph{Monotonicity}~\citep{Arya2019One} measures the increase in classification probability when inserting pixels {on a black background after normalization}. 
However, instead of evaluating the probability directly, it assesses whether the classification probability increases monotonically as pixels are inserted according to their saliency values.
High values in Insertion and Monotonicity as well as a low value in Deletion indicate that the saliency map captures the most important parts of the input. 

{As a metric for robustness, \emph{Avg.Sensitivity}~\citep{Yeh2019Fidelity} quantifies how much a saliency map changes in response to small input perturbations. Specifically, perturbations are sampled from a
fixed radius around the original input, and the average difference between the saliency maps of the perturbed and original inputs is computed. A higher Avg.Sensitivity indicates that the saliency map is more sensitive to
small perturbations and therefore less robust.}

Regarding complexity, 
\emph{Ent.Complexity} \citep{Bhatt2020Evaluating} evaluates the entropy complexity, that is, measuring the entropy of the fractional contribution distribution of the input element.
\emph{Sparseness}~\citep{Chalasani2020Concise} employs the Gini Index to assess whether the model's prediction relies primarily on a few highly attributed features. A lower \emph{Ent.Complexity} value and a higher \emph{Sparseness} value indicate a more concise saliency map.

Finally, for localization, 
\emph{Energy Pointing Game} (EPG; \citep{Wang2020Score-CAM}) measures the fraction of the saliency map's mass that falls within the union of the ground truth bounding boxes. A higher value signifies improved performance
on localization.

\subsubsection{User Reactions and Abilities}

Unless stated otherwise, participants rated all items on a five-point scale ranging from 1 (strongly disagree) to 5 (strongly agree). We measured explanation satisfaction with 8 items from \citet{Hoffman2018Metrics}, \eg
\enquote{the saliency maps are useful to evaluate the classifier} {(Cronbach's $\alpha = .80$)}. Trust in the saliency maps {($\alpha = .81$)} and the classifier {($\alpha = .84$)} were measured with
three items respectively, using the same scale from \citet{thielsch2018trust}, but referring to either the saliency maps or the classifier. A sample item is \enquote{I would trust the saliency maps/the classifier
completely.} Classifier trustworthiness was measured using 10 items {($\alpha = .89$)}: one self-constructed item, \ie \enquote{I think the image classifier is trustworthy}, and others from the trustworthiness subscales
ability and transparency from \citet{hoeddinghaus2021automation}. The subscale ability consists of six items; a sample item is \enquote{I think the classifier has the competence to adapt its decision to different
circumstances} {($\alpha = .87$)}. The subscale transparency consists of three items; a sample item is \enquote{I think the decision-making processes of the classifier are clear and transparent}
{($\alpha = .84$)}. 

For abilities, participants in each trial were first asked to estimate the accuracy of the classifier for the respective class from 0 to 100\%. We then calculated a variable that we named \enquote{absolute accuracy difference.} It reflects the absolute differences between the real classification accuracy and the human-estimated accuracy for each trial. Higher values thus mean a stronger deviation from the real accuracy. The real classification accuracy for each class was calculated as follows: For each image, the class with the highest probability was taken as the predicted class. The number of correctly classified images, that is, whether the model prediction aligns with the human annotation in the training data, was then divided by the total number of images for each class.

Participants then judged which of the 6 additional images of the respective class would be correctly or incorrectly classified based on the three saliency maps of different images they saw before. For each round, we counted prediction errors, that is, cases in which participants' judgments did not match the classifier's actual output. The participants' \enquote{error rate} was calculated by dividing the number of errors by the total number of predictions.
\section{Results}

We used ANOVA, \textit{t}-tests, and linear mixed models (LMM) to analyze the data. For the absolute accuracy difference and prediction error rate (except in the LMM analyses), mean values of the 12 trials for each participant were first calculated. We assessed the assumptions of homogeneity of variance and normality using Levene's tests, Shapiro--Wilk tests, and visual inspection of Q--Q plots. The assumptions were met except for the analyses involving absolute accuracy difference and prediction error rate, where deviations from normality were observed; therefore, these results were interpreted with caution.

\subsection{User Reactions}

\textbf{RQ1.1} asked whether perceived explanation satisfaction differed across groups. A Type~III ANOVA revealed no significant differences between saliency method conditions, $F(2, 163) = 0.48$, $p = .618$. Estimated marginal means for explanation satisfaction were comparable across groups: Grad-CAM ($M = 3.98$, $SE = 0.09$), GBP ($M = 3.92$, $SE = 0.08$), and LIME ($M = 3.87$, $SE = 0.07$). None of the pairwise comparisons were statistically significant: Grad-CAM vs. LIME, $t(163) = 0.98$, $p = .33$, $d = 0.19$; Grad-CAM vs. GBP, $t(163) = 0.56$, $p = .58$, $d = 0.11$; GBP vs. LIME, $t(163) = 0.39$, $p = .70$, $d = 0.07$. Thus, in response to \textbf{RQ1.1}, explanation satisfaction did not significantly differ between the conditions.

\textbf{RQ1.2} asked whether trust in the classifier and in the saliency map differs between the explanation groups. To test this, we ran three separate ANOVAs for trust in the classifier, trust in the saliency map, and perceived trustworthiness of the classifier. For overall trustworthiness of the classifier, the ANOVA was not significant, $F(2, 163) = 1.18$, $p = .31$, with post hoc comparisons revealing no significant pairwise differences (all $p > .13$). Similarly, for trust in the classifier, no significant group differences were found, $F(2, 163) = 0.10$, $p = .91$. All pairwise comparisons were non-significant ($p > .67$). Finally, trust in the saliency map also did not differ significantly between groups, $F(2, 163) = 0.14$, $p = .87$. Post hoc tests again revealed no significant pairwise differences ($p > .63$). The answer to \textbf{RQ1.2} is therefore that no significant differences in trust-related perceptions emerged between the saliency map groups.

\subsection{User Abilities}

\autoref{tab:accrucy_by_trail} shows the real classifier accuracy, the mean estimated accuracy, and the standard deviation for each of the 12 classes. Although participants' mean accuracy estimations over all trials and participants are close to the real overall classifier accuracy, the standard deviation is substantial. This indicates that participants struggled to assess classifier accuracy using the saliency maps provided.

\begin{table*}[!htbp]
    \centering
    \setlength{\tabcolsep}{5pt}
    \begin{tabular}{c ccccccccccccc}
        \toprule
         & 1 & 2 & 3 & 4 & 5 & 6 & 7 & 8 & 9 & 10 & 11 & 12 & Overall \\
        \midrule
        Classifier & 54 & 92 & 60 & 84 & 52 & 88 & 54 & 76 & 76 & 54 & 68 & 86 & 70 \\
        Participants (mean) & 73 & 71 & 73 & 61 & 68 & 68 & 73 & 67 & 56 & 69 & 72 & 58 & 67 \\
        Participants (standard deviation) & 12 & 13 & 14 & 18 & 14 & 14 & 12 & 17 & 19 & 18 & 16 & 19 & 15 \\
        \bottomrule
    \end{tabular}
    \caption{Real accuracy percentages of classifier and estimated accuracy by participants by trial.}
    \label{tab:accrucy_by_trail}
\end{table*}

\textbf{RQ2.1} asked whether participants' accuracy predictions differed from the true accuracy depending on the saliency map. A Type~III ANOVA revealed a significant main effect of the map on the deviation between predicted and actual accuracy, $F(2, 163) = 5.69$, $p = .004$. Planned contrasts using estimated marginal means comparisons showed that Grad-CAM led to significantly smaller prediction errors than both GBP ($t(163) = -2.13$, $p = .034$, $d = -.43$) and LIME ($t(163) = -3.36$, $p = .001$, $d = -.63$). No significant difference was found between GBP and LIME ($t(163) = -1.09$, $p = .279$, $d = -.20$).\footnote{Because residual normality was violated for the dependent variable \enquote{absolute accuracy difference}, we additionally conducted non-parametric Kruskal--Wallis tests with Holm-corrected pairwise post-hoc comparisons. The pattern of statistically significant effects remained unchanged.} The response to \textbf{RQ2.1} is therefore that participants using Grad-CAM showed more accurate predictions than those using LIME and GBP.

\textbf{RQ2.2} asked whether groups differed in their error rates when predicting classifier correctness. A Type~III ANOVA revealed a significant main effect of saliency method on error rate, $F(2, 163) = 3.76$, $p = .013$. Estimated marginal means indicated that participants in the Grad-CAM group ($M = 0.375$, $SE = 0.009$) had a significantly lower error rate than those in the LIME group ($M = 0.408$, $SE = 0.008$), $t(163) = -2.74$, $p = .007$, $d = -0.52$. The Grad-CAM group also outperformed the GBP group ($M = 0.395$, $SE = 0.009$), but this difference did not reach statistical significance, $t(163) = -1.58$, $p = .117$, $d = -0.32$. There was no significant difference in error rate between the GBP and LIME groups, $t(163) = -1.07$, $p = .288$, $d = -0.20$. \footnote{Because residual normality was violated for the prediction error rate, we additionally conducted non-parametric Kruskal--Wallis tests with Holm-corrected pairwise post-hoc comparisons. The pattern of statistically significant effects remained unchanged.} The response to \textbf{RQ2.2} is therefore that Grad-CAM led to fewer errors when predicting classifier correctness than LIME, but not significantly fewer than GBP.

\subsection{Mathematical Metrics}

\textbf{RQ3} asked whether mathematical metrics differ across the three approaches. 
We assessed the mathematical properties using the metrics discussed in \autoref{sec:mat:metrics} and calculated contrasts using estimated marginal means comparisons. The results are shown in \autoref{tab:mean_properties}. 
Across the seven evaluated metrics, GBP showed the most consistently favorable mathematical properties. 
Specifically, GBP achieved the most favorable mean values compared to Grad-CAM and LIME in three metrics (Deletion, Sparseness, and Ent.Complexity). Grad-CAM and GBP jointly achieved the most favorable values in Monotonicity.
Although the pairwise comparison for Monotonicity between Grad-CAM and GBP did not reach statistical significance, the estimated effect size was large and in favor of GBP (Cohen's $d = 2.1$), suggesting a potentially practically relevant difference that may not have been detected due to limited statistical power (only 12 observations per metric and approach). 
Grad-CAM and LIME additionally shared the most favorable values in Avg.Sensitivity. 
LIME did not exhibit the most favorable values in any metric when compared to both GBP and Grad-CAM.

Taken together, these results indicate that GBP demonstrated more favorable mathematical properties than both alternative approaches across a greater number of metrics, whereas Grad-CAM showed strengths in selected metrics but less consistently so. 
Thus, in response to \textbf{RQ3}, GBP can be considered superior to LIME and Grad-CAM in terms of mathematical metrics, with Grad-CAM ranking second and LIME showing the weakest overall performance.

\begin{table*}[h!]
\centering
\setlength{\tabcolsep}{2pt}
\begin{tabular}{l ccc cc c c}
\toprule
\textbf{Group} & \multicolumn{3}{c}{Fidelity} & \multicolumn{2}{c}{Complexity} & Robustness & Localization \\\cmidrule(lr){2-4} \cmidrule(lr){5-6} \cmidrule(lr){7-7} \cmidrule(lr){8-8} 
&{Insertion} $\uparrow$ & {Deletion}  $\downarrow$  & {Monotonicity} $\uparrow$ & {Sparseness} $\uparrow$ & {Ent.Complexity} $\downarrow$ & {Avg. Sensitivity}  $\downarrow$ & {EPG}  $\uparrow$\\
\midrule
\textbf{Grad-CAM} & 0.9013 & 0.1273            & \textbf{0.7341*}          & 0.4283          & {10.4929} & \textbf{0.9102*}          & 0.39\\
\textbf{GBP}     & 0.8504          & \textbf{0.0687*}   & \textbf{0.8299*} & \textbf{0.8081*} & \textbf{9.2976*}          & {1.6733} & 0.44\\
\textbf{LIME}    & 0.8744          & 0.1277            & 0.4857          & 0.5564          & 10.1132          &\textbf{0.9012*}          & 0.41\\
\bottomrule
\end{tabular}
\caption{Answering RQ3. Mean values for the mathematical properties in the 12 trials across the different groups (Grad-CAM, GBP, LIME). An upward arrow $\uparrow$ indicates that a higher value is preferable, while a downward arrow $\downarrow$ signifies that a lower value is better. Bold values and * indicate the most favorable values for each property with p < .05. More than one bolded value in a single column means a shared first place. The corresponding standard deviations can be found in Appendix C.}
\label{tab:mean_properties}
\end{table*}

To assess \textbf{RQ4}, whether the mathematical properties of the saliency maps are associated with accuracy predictions, we used the mean of each metric for the three saliency maps in each trial. 
We then calculated one LMM per saliency map group for each of the seven metrics. 
For each LMM, we introduce a dependent variable called \enquote{absolute accuracy difference}, which is the absolute differences between participants' accuracy ratings and the true accuracy for each specific class.
To account for the repeated-measures structure of the data, we included a random intercept for participants. This random intercept accounted for individual differences in baseline accuracy estimation, as each participant provided predictions across all twelve trials.  For example, in the model assessing the influence of the \enquote{insertion} metric, we specified an LMM with
\enquote{absolute accuracy difference} as the dependent variable, \enquote{insertion} as the fixed effect, and random intercepts for participants.\footnote{Although residuals showed deviations from normality for models predicting
\enquote{absolute accuracy difference}, we retained the LMM approach because linear mixed models are generally robust to moderate violations of distributional assumptions, particularly in studies with repeated
observations and random-effects structures~\citep{schielzeth2020robustness}. Nevertheless, these results should be interpreted with appropriate caution.}

The LMM results can be seen in \autoref{tab:results}.\footnote{More detailed results, including standardized effect estimates ($\beta$), 95\% confidence intervals, and significance values for all tested mathematical metrics and saliency map groups, are provided in Supplementary Table~6.}
A negative value indicates that higher values in the metric are associated with smaller absolute accuracy differences, whereas a positive value indicates that higher values in the metric are associated with larger absolute accuracy differences.
Reducing the absolute accuracy difference means decreasing the deviation between the human prediction of accuracy and the actual accuracy based on saliency maps. Thus, if the properties evaluated by the mathematical metric contribute to human accuracy prediction, we expect positive values for \enquote{Deletion,} \enquote{Ent.Complexity,} and \enquote{Avg.Sensitivity,} and negative values for \enquote{Insertion,} \enquote{Monotonicity,} \enquote{Sparseness,} and \enquote{EPG}. 
However, our analysis yielded mixed results.

Significant results in the expected direction were found for \enquote{Deletion} and \enquote{Sparseness} in the GBP and LIME groups, for \enquote{Ent.Complexity} in the GBP group, and for \enquote{EPG} in the Grad-CAM and GBP groups, indicating these metrics were linked to better accuracy predictions in those groups. 
Significant results in the unexpected direction were found for \enquote{Insertion} in the LIME group, for \enquote{Monotonicity} in the GBP group, for \enquote{Ent.Complexity} in the Grad-CAM group, and for \enquote{Avg.Sensitivity} in the Grad-CAM and LIME groups. This implies that improvements of these metrics in the specific groups were associated with worse accuracy predictions. The response to \textbf{RQ4} thus is that while there are associations between mathematical metrics and accuracy predictions, the picture is complex. Many of the mathematical metrics are inversely related to human understanding, and \enquote{Deletion,} \enquote{Sparseness,} and \enquote{EPG} were the only metrics for which all significant associations were in the expected direction.

\begin{table*}[h!]
\centering
\setlength{\tabcolsep}{2pt}
\begin{tabular}{l ccc cc c c} \toprule
\textbf{Group} & \multicolumn{3}{c}{Fidelity} & \multicolumn{2}{c}{Complexity} & Robustness & Localization \\\cmidrule(lr){2-4} \cmidrule(lr){5-6} \cmidrule(lr){7-7} \cmidrule(lr){8-8} 
&{Insertion} $\uparrow$ & {Deletion}  $\downarrow$  & {Monotonicity} $\uparrow$ & {{Sparseness} $\uparrow$} & {Ent.Complexity} $\downarrow$ & {Avg. Sensitivity}  $\downarrow$ & {EPG}  $\uparrow$\\
\midrule
\textbf{Grad-CAM} & -6.85           & 2.36             & -5.53           & 13.97            & \textbf{-15.26*} & \textbf{-305.60*} & \textbf{-13.80*}\\
\textbf{GBP}     & 0.54            & \textbf{43.43*}  & \textbf{32.37*} & \textbf{-46.33*} & \textbf{9.03*}   & -1.63             & \textbf{-17.78*}\\
\textbf{LIME}    & \textbf{9.54*}  & \textbf{24.32*}  & -0.86           & \textbf{-13.33*} & -1.38            & \textbf{-95.04*}  & -5.38\\
\bottomrule
\end{tabular}
\caption{Answering RQ4. Table showing linear mixed model results for the impact of different mathematical properties on the accuracy difference ratings for each saliency map approach. The analysis for each metric was conducted separately. * indicates \textit{p} < .05. \textit{N} = 166}
\label{tab:results}
\end{table*}

\section{Discussion}

In this study, we aimed to assess the alignment between user and mathematical metrics using popular saliency map approaches. Our main findings show that, even though GBP had an edge over Grad-CAM in overall mathematical metrics, Grad-CAM led to the best task performance, indicating a misalignment between mathematical metrics and user abilities (see \autoref{tab:rq-summary}).

\newcolumntype{Y}{>{\raggedright\arraybackslash}X}

\begin{table*}[t]
\centering
\caption{{Summary of findings by research question.}}
\label{tab:rq-summary}
\footnotesize
\renewcommand{\arraystretch}{1.18}
\setlength{\tabcolsep}{5pt}

\begin{tabularx}{\textwidth}{
    >{\raggedright\arraybackslash}p{0.07\textwidth}
    >{\raggedright\arraybackslash}p{0.17\textwidth}
    Y
    >{\raggedright\arraybackslash}p{0.40\textwidth}
}
\toprule
\textbf{RQ} &
\textbf{Evaluation family} &
\textbf{Outcome examined} &
\textbf{Main finding} \\
\midrule

RQ1.1 &
Subjective &
Perceived explanation satisfaction &
No significant differences were found for perceived explanation satisfaction. The three approaches were perceived as similarly satisfactory. \\

\addlinespace

RQ1.2 &
Subjective &
Trust in the classifier, trust in the saliency maps, and perceived
classifier trustworthiness &
No significant differences were found for trust-related measures. The three approaches performed similarly across trust-related measures. \\

\addlinespace

RQ2.1 &
Objective &
Difference between participants' estimated classifier accuracy and
the classifier's actual accuracy &
Grad-CAM supported participants in estimating the classifier's
accuracy significantly better than both GBP and LIME. \\

\addlinespace

RQ2.2 &
Objective &
Errors when predicting which individual images would be classified
correctly &
Grad-CAM supported participants in making classification predictions significantly better than
LIME, but not significantly better than GBP. \\

\addlinespace

RQ3 &
Mathematical &
Metrics of fidelity, complexity, robustness, and localization &
GBP showed the most consistently favorable mathematical results, followed by Grad-CAM. LIME did not outperform both alternatives on any metric.
\\

\addlinespace

RQ4 &
Relationship between mathematical and objective evaluation &
Associations between mathematical metrics and participants' accuracy-estimation errors &
Some relationships were in the expected direction, whereas several others were counterintuitive. Mathematical metrics were not  straightforward proxies for
objective user performance. \\

\midrule
\multicolumn{4}{p{\dimexpr\textwidth-2\tabcolsep\relax}}{
\textbf{Overall finding.}
The three evaluation families led to different conclusions: subjective measures revealed no meaningful differences, objective performance favored Grad-CAM, and mathematical metrics favored GBP. The results therefore indicate a misalignment between subjective, objective, and mathematical XAI evaluation.
} \\
\bottomrule
\end{tabularx}
\end{table*}

\subsection{Contextualizing the Results}

\subsubsection{Performance across Evaluation Methods}

All three approaches performed roughly equally regarding trust and satisfaction ratings. 
This is in contrast to findings by \citet{Khadivpour2022Responsibility}, where participants preferred Grad-CAM and rated it as more understandable than LIME. 
This discrepancy could be caused by differences in measurement, as participants in the study by \citet{Khadivpour2022Responsibility} simply ranked their preference while we employed the 8 items from the explanation
satisfaction scale (which might be a slightly different construct as it focuses more on usefulness, see \citep{Hoffman2018Metrics}).

{When looking at ability-related measures of understanding, our results are broadly consistent with \citet{morrison2023eye}, who found that Grad-CAM was more helpful for users than LIME in a similar task.
Likewise, several studies comparing Grad-CAM with other saliency approaches reported advantages for Grad-CAM on at least one performance measure, although the overall pattern of results was mixed rather than consistently
favoring a single approach \citep{Kim2022HIVE,Colin2022What}. Consistent with this literature, our findings provide evidence for advantages of Grad-CAM on some, but not all, comparisons.}

{One possible explanation} is that Grad-CAM tends to hit a human-usable semantic granularity: specific enough to indicate the relevant region, but not so detailed that it overwhelms or invites overinterpretation. The other two approaches can both miss the level that best supports human judgment: LIME can be too coarse (due to large superpixels) or oddly shaped; GBP can be too fine due to its focus on edges.

Regarding mathematical metrics, our findings are consistent with previous observations by \citet{zhang2024saliency} and \citet{li2021experimental} who also found that no single saliency method dominates across all metrics.
Conversely, GBP showing the best performance across metrics contrasts with these studies, where Grad-CAM performed best. 

This difference might arise from the dataset scope and the metrics selected. Specifically, both works choose metrics suited to their respective objectives, which include a range of metrics that favor or disadvantage certain saliency maps by design. 
We focus on a few widely recognized metrics rather than benchmarking or analyzing metric impacts.
Furthermore, when evaluating one property using multiple metrics, we also observe disagreements between them (similar to \citep{zhang2024saliency}).
Overall, these results make it challenging to determine which saliency map was superior based on mathematical evaluations.

\subsubsection{Mathematical Metrics and User Predictions}

The association between mathematical metrics and the accuracy ratings was complex and varied between metrics. Furthermore, because these analyses were based on a comparatively small number of classes, the observed associations should be interpreted cautiously.
For fidelity, we expected that saliency maps that more accurately mirror the model's underlying dynamics (i.e.,~have higher scores on fidelity metrics) would lead to higher user understanding. {Fidelity
metrics are intuitively appealing because they assess whether highlighted pixels genuinely influence the model's prediction. By measuring prediction changes after adding or removing relevant pixels, they operationalize a
counterfactual notion of feature importance that is conceptually related to human judgments of evidential relevance~\citep{petsiuk2018rise}.}
We observed that Deletion was associated with the absolute accuracy difference as expected, but this was not the case for Insertion and Monotonicity. In fact, we found counterintuitive effects in the LIME group for Insertion and in the GBP group for Monotonicity. 
If the goal is to best approximate user predictions, {our results provide evidence that Deletion may better approximate user predictions than the other fidelity metrics considered here, although this finding should be validated in larger datasets.}

The results for insertion and monotonicity may emerge because these metrics are highly sensitive to how the baseline is constructed (e.g., through blur, mean, or noise) and how pixels are re-introduced (especially regarding the resulting out-of-distribution intermediate images). There may be artifacts that rapidly increase confidence, even when these are not explanatory for humans. By contrast, deletion more directly tests the impact of pixels, which may explain why it aligns more consistently with users' accuracy predictions in our study.

A low complexity is expected to aid humans in understanding saliency maps~\citep{nauta2023anecdotal}. 
From an intuitive perspective, low complexity, like Sparseness, could improve user performance by reducing visual clutter and directing attention to a few relevant features, thereby possibly lowering cognitive
load~\citep{das2024shifting}. However, its benefit can depend on whether the highlighted features remain informative and meaningful for the task~\citep{fox2024cognitive}.
The experimental results suggest that the role of complexity is saliency-method dependent. For GBP, both Sparseness and Ent.Complexity consistently indicate that concise and focused explanations facilitate accurate human predictions. For LIME, this effect is only partially supported by Sparseness. In contrast, Grad-CAM exhibits the opposite pattern, where greater Ent.Complexity is associated with more accurate human judgments, suggesting that users may benefit from richer visual information when interpreting Grad-CAM explanations.

Against expectations \citep{nauta2023anecdotal}, our results suggest that lower complexity does not necessarily promote better human understanding. The chosen complexity metrics quantify how concentrated or dispersed attribution mass is, but humans may
benefit the most from maps that are not only sparse but also semantically well-structured and task-relevant. In other words, the relationship between complexity and understanding is plausibly non-monotonic: while overly
complicated saliency maps certainly overwhelm, too sparse saliency maps may end up having too little context, thus hindering comprehension.

{Higher robustness may help users by making the model's reactions more visible, clarifying which features influence predictions.}
Our results show that higher Sensitivity is associated with better accuracy ratings, suggesting that humans tend to benefit from more sensitive saliency map methods. This is contrary to our (and common) expectations, as we predicted less sensitive maps to deliver more reliable insight into the functioning of the classifier. One explanation for this could be that higher Sensitivity enables humans to better recognize what features the classifier reacts to the most.

{Localization aligns with human perception because it measures overlap with object regions, matching how humans interpret images at the object level.}
EPG, a metric for evaluating localization, measures the overlap between the saliency map and the human-annotated object using a bounding box, aligning with human annotation of objects by concept. This alignment may explain the positive relation between EPG and human understanding. Of the metrics we examined, EPG seems to be the one most closely associated with user accuracy predictions, at least in the GBP and Grad-CAM groups. This means that localization might be the most useful to serve as a proxy for user metrics.

These results provide more nuance to the common hypothesis that mathematical metrics can serve as a proxy for human studies \citep{DoshiVelez2017Towards,carvalho2019machine}. Not only did the different approaches fare
differently across the saliency map methods, but the connection between the mathematical metrics and the accuracy rating abilities measure were also rather mixed. According to our results, Deletion, Sparseness, and EPG could provide
an indication of how users might fare in terms of abilities and even then, only for specific saliency map approaches. Interestingly, this finding is in contrast to findings by \citet{Colin2022What}, who found that EPG
was not associated with indicating user performance. As things stand, this means that using the mathematical metrics examined in the current study as proxies for findings of user studies should therefore be done cautiously, particularly for saliency maps in image classification tasks.

\subsection{Practical Implications}

\subsubsection{Implications for Using Saliency Maps}

Our study does not identify a best saliency map method, suggesting that the optimal choice depends on the specific context. If users' abilities and objective understanding are a priority, Grad-CAM appears to be the most
suitable option. Both our findings and prior literature (e.g.,~\citep{morrison2023eye}) indicate that Grad-CAM helps users understand classifiers. Additionally, our results demonstrate that it fosters a moderate level of
trust and a high degree of explanation satisfaction. On the other hand, in contexts where faithfulness to the underlying model is paramount---such as regulatory audits or safety-critical applications---our findings suggest
GBP may be a more appropriate choice. However, as some studies have also identified Grad-CAM as superior in this regard \citep{li2021experimental}, the evidence remains inconclusive, and further research is needed to provide a definitive recommendation.

Our results are also in line with research suggesting that saliency maps may have limited usefulness in helping participants make accurate classification predictions (e.g.,~\citep{Alqaraawi2020Evaluating}). In our
experiment, participants' average accuracy predictions deviated by $\pm$18--21\% from the actual accuracy, depending on the saliency group (18.1\% for GradCAM, 20.3\% for GBP, 21.3\% for LIME). The classification prediction error rate was 37\%--40\%, only slightly better chance level. This is particularly
noteworthy because we only examined participants with a background in ICT in our study. As these participants likely have a better understanding of saliency maps than laypeople,
objective measures of user understanding might be even worse for a broader audience.

\subsubsection{Implications for {{XAI}} Evaluation}

The finding that the evaluation methods used in our study diverged in their assessment raises fundamental questions about the alignment between mathematical and user evaluations in XAI. 
While mathematical metrics are crucial to ensure that explanations are technically sound and accurately reflect the inner workings of a model, our results indicate that they cannot be easily associated with user abilities. 
This discrepancy suggests that focusing solely on mathematical metrics may overlook critical aspects of how users interact with and benefit from explanations. 
Instead, our findings underscore the need for a holistic assessment of user and mathematical metrics. For saliency maps, for example, this could be a structured method of classification such as saliency or property cards
\citep{Boggust2023Saliency,Speith2026Review}. For XAI evaluation in general, we need multi-modal evaluation toolkits (see, e.g.,~\citep{Speith2023New}). 
{This is consistent with broader approaches to AI governance, which emphasize that technical mechanisms should be integrated with human-centered evaluation and institutional oversight \citep{Odion2026AI}.}

Furthermore, the trust and satisfaction ratings of users did not provide much insight into the meaningfulness and usefulness of an explainability approach for user abilities, and thus their actual understanding of a classifier. 
This aligns with findings in cognitive science (e.g.,~\citep{Rozenblit2002Misunderstood}), demonstrating that perceived understanding often diverges from actual understanding.
Such subjective measures can be influenced by cognitive biases, user familiarity with the explanation style, or even the aesthetic appeal of an XAI approach's outputs and might not translate into increased understanding
and the factual correctness of the approach \citep{bertrand2022cognitive}. 
While subjective user ratings can still be important for factors like adoption, examining user abilities and mathematical properties are important complements to mitigate such risks, especially as our study clearly shows that human reactions, user abilities, and mathematical metrics can point in different directions. Overall, our view is that a meaningful XAI evaluation should ideally cover all modalities.

\subsection{Limitations and Future Work}

One limitation of our study is that we only had English-speaking participants with a background in ICT. This limits the sample in terms of culture and expertise. In particular, it is likely that lay people who are confronted with saliency maps have a different attitude towards these explainability methods. However, we deliberately chose these participants, as they are the most likely to encounter and use saliency maps in their professional work, and thus possess the necessary expertise to critically evaluate them. 
Future studies should follow a similar procedure with other stakeholder groups.

Furthermore, we limited our study to only one classifier. Other studies with multiple classifiers have found differences in user performance using different classifiers \citep{Colin2022What,Nguyen2021Effectiveness}. While we
encourage future research to extend our design to multiple classifiers, our limited design already highlights the most important issue: we cannot assume that the different evaluation methods are aligned. This confirms and
reinforces initial findings from earlier settings \citep{Colin2022What,Nguyen2021Effectiveness}. This also mitigates the limitation of only using saliency maps: while future research should clearly also examine other
explainability approaches, we believe that saliency maps were a suitable testbed for drawing attention to the divergence of evaluations. Further research must now systematize this across populations, classifications, and explainability approaches.

{A further limitation concerns the number of statistical tests conducted throughout this study, which generally increases the risk of Type~I errors. This issue is particularly pertinent for the analyses addressing RQ3 and RQ4, as they comprise multiple statistical tests across several outcomes and mathematical metrics. Moreover, the analyses relating mathematical metrics to participant performance (RQ4) were conducted on only 12 image classes, limiting the precision of the estimated associations and increasing the possibility that weaker relationships remained undetected. Consequently, findings from RQ3 and RQ4 should be interpreted with appropriate caution and validated in future studies using a larger and more diverse set of image classes.}

{Another limitation is that our study was conducted in a low-risk ImageNet-based image classification setting, where incorrect classifications have no direct consequences for participants. Trust in such settings may differ from trust in high-stakes domains, where users must consider potential risks, accountability, and consequences of AI errors. Therefore, participants' trust judgments in our study may reflect a more heuristic evaluation of the explanations rather than the deeper reliance that can occur when AI decisions affect important outcomes. This lower level of perceived risk may also have reduced the salience of trust differences between saliency map approaches, potentially contributing to the absence of differences in trust-related measures. Future research should investigate whether the relative influence of different explanation approaches changes in high-stakes domains, such as healthcare or cybersecurity, where users have stronger incentives to scrutinize and rely on AI explanations.}

Lastly, we presented the Grad-CAM saliency maps with the original image clearly visible in the background, whereas for LIME, the original image was only faintly visible, and for GBP, no background image was provided. This
is a common presentation for each of these approaches \citep{Adebayo2018Sanity,Selvaraju2017Grad-CAM,Khadivpour2022Responsibility,Kierdorf2023Reliability}. Nonetheless, having the original image in the background of the
saliency map may assist users by reducing the cognitive demands to mentally overlay the original with the saliency map to identify which features contributed to the classification. This could partly explain why
participants in the Grad-CAM group demonstrated the highest user abilities compared to other approaches, though the finding is still consistent with prior research (e.g.,~\citep{morrison2023eye, Kim2022HIVE,
Colin2022What}). To reduce this discrepancy, we displayed the original image directly above the corresponding saliency map for all groups.

\section{Conclusion}

This study used saliency maps as a testbed to examine how different families of evaluation metrics in \ac{XAI}---sub\-ject\-ive user judgments, objective user performance, and mathematical metrics---align or diverge. 
While participants reported no {significant} differences in trust or satisfaction across Grad-CAM, LIME, and GBP, Grad-CAM most effectively supported users in predicting model performance, whereas GBP achieved the most favorable ratings on mathematical metrics. 
Importantly, the association between mathematical metrics and user understanding was not straightforward and often counterintuitive. 
These findings highlight a broader challenge in \ac{XAI}: subjective, objective, and mathematical evaluations do not necessarily converge, and technically favorable metrics do not guarantee better user outcomes. 
Rather than treating mathematical metrics as substitutes for user studies, our results suggest that they should be seen as complementary, each capturing different dimensions of explanation quality. 
Future work should investigate how to systematically combine metric families in order to meet context-dependent desiderata of \ac{XAI} approaches.

\section*{Ethics Statement}

The authors confirm that this work complies with ethical standards. All necessary institutional approvals have been obtained where applicable.

\section*{Generative AI Disclosure Statement}
We used ChatGPT and DeepL to improve the grammar and fluency of writing. 

\section*{Funding Statement}

This work was supported by:

\begin{itemize}
\item {Volkswagen Foundation}[http://dx.doi.org/10.13039/501100001663] (Grant No. {AZ9B830}, {AZ98513}, {AZ98514}), as part of the project \textit{Explainable Intelligent Systems (EIS)}
\item {Deutsche Forschungsgemeinschaft (DFG)}[http://dx.doi.org/10.13039/501100001659], Grant No. {389792660}, as part of TRR 248 \textit{CPEC -- Center for
Perspicuous Computing}
\item {Daimler and Benz Foundation}[http://dx.doi.org/10.13039/501100002974], project \textit{TITAN -- Technologische Intelligenz zur Transformation,
Automatisierung und Nutzerorientierung des Justizsystems} (Grant No. {45-06/24})
\end{itemize}

\bibliographystyle{ACM-Reference-Format}
\bibliography{bibliography}

\appendix

\section{Survey Material}
\label{saliency::app::survey}

\subsection{Introduction}
\label{saliency::app::intro}

Welcome to the Study!

Thank you for being interested in our experiment. This research is \textit{[redacted for review]}.

In this experiment, you will evaluate an image classifier based on AI. The classifier uses a visualization technique called \enquote{saliency map}. Don't worry if you have not heard of this before, all important terms will be explained at the beginning of the study.

Please only take part in this study if you are over the age of 18, proficient with the English language and on a PC, laptop or (large) tablet. The experiment takes around 40 minutes.

\begin{enumerate}
\setcounter{enumi}{\value{qcounter}}
    \item Please insert your Prolific ID here \emph{[free text]}
\setcounter{qcounter}{\value{enumi}}
\end{enumerate}

\subsection{Data Protection}

\textbf{Consent according to the EU General Data Protection Regulation (GDPR)}

\noindent\textbf{1. What personal data do we collect and process?}

\noindent We take the protection of your personal data very seriously. We treat your personal data confidentially and in accordance with the statutory data protection regulations and this privacy policy. Participation in the study is anonymous, and no conclusions can be drawn about your identity under any circumstances.

We use your personal data for conducting our study in compliance with the applicable data protection provisions. Below, we will explain which personal data is collected and stored. You will also receive information on how your data is used and what rights you have regarding the use of your data.

The following personal data is collected when participating in the study:

\begin{itemize}
    \item Gender
    \item Age
    \item Education
    \item Experience with Saliency Maps
\end{itemize}

This data is collected exclusively for scientific purposes. By signing this document, you permit \textit{[redacted for review]} to use the information you provide in the survey for these scientific purposes.\\

\noindent\textbf{2. Information about the rights of data subjects}

\noindent You have the right to:

\begin{itemize}
    \item Request information about your personal data processed by us in accordance with Art. 15 GDPR. In particular, you can request information about the processing purposes, the category of personal data, the categories of recipients to whom your data has been or will be disclosed, the planned storage duration, the existence of a right to rectification, deletion, restriction of processing or objection, the existence of a right to complain, the origin of your data if it was not collected by us, and the existence of automated decision-making including profiling and, if applicable, meaningful information about its details.
    \item Request the correction of incorrect or incomplete personal data stored by us without delay in accordance with Art. 16 GDPR.
    \item Request the deletion of your personal data stored by us in accordance with Art. 17 GDPR, unless the processing is necessary for exercising the right to freedom of expression and information, to fulfill a legal obligation, for reasons of public interest, or to assert, exercise, or defend legal claims.
    \item Request the restriction of processing of your personal data in accordance with Art. 18 GDPR, as far as the accuracy of the data is disputed by you, the processing is unlawful, but you oppose its deletion, and we no longer need the data, but you require it to assert, exercise, or defend legal claims or you have objected to the processing in accordance with Art. 21 GDPR.
    \item Receive your personal data that you have provided to us in a structured, commonly used, and machine-readable format or to request the transfer to another controller in accordance with Art. 20 GDPR.
    \item Withdraw your consent at any time in accordance with Art. 7 para. 3 GDPR. This means that we will no longer continue the data processing based on this consent in the future.
    \item Complain to a supervisory authority in accordance with Art. 77 GDPR. As a rule, you can contact the supervisory authority of your usual place of residence or workplace or our office.\\
\end{itemize}

\noindent\textbf{3. Information about the right to object}

\noindent If your personal data is processed based on legitimate interests pursuant to Art. 6 para. 1 sentence 1 lit. f GDPR, you have the right to object to the processing of your personal data in accordance with Art. 21 GDPR, provided that there are reasons arising from your particular situation or the objection is directed against direct marketing. In the latter case, you have a general right to object that we will implement without the need to specify a particular situation.

If you wish to exercise your right of withdrawal, an email to \textit{[redacted for review]} is sufficient.

The supervisory authority responsible for \textit{[redacted for review]} is \textit{[redacted for review]}

Data retention period: Seven years.

\begin{enumerate}
\setcounter{enumi}{\value{qcounter}}
        \item Do you consent to these terms? 
            \mychoice{Yes, I consent.}
            \mychoice{No, I do not consent (end experiment)}
\setcounter{qcounter}{\value{enumi}}
\end{enumerate}

Please read the explanations on the following pages carefully. You will be asked to answer comprehension questions later, and you cannot take part in the study if you answer incorrectly twice in a row.

If you are already proficient with the concept of evaluating AI image classifiers using saliency maps and confident you can answer comprehension questions, you may skip the explanations.

\subsection{Background Information}
\label{app:background}

\textbf{Please read the explanations on the following pages carefully. You will be asked to answer comprehension questions later, and you cannot take part in the study if you answer incorrectly twice in a row.}

\textbf{If you are already proficient with the concept of evaluating AI image classifiers using saliency maps and confident you can answer comprehension questions, you may skip the explanations.}\\

\noindent\textbf{What are Saliency Maps}

\noindent In this study, we will be exploring something called \enquote{saliency maps.} To help you understand what these are, let's start with a simple explanation.

Imagine you are looking at a picture. Your eyes and brain naturally focus on certain parts of the image more than others. For example, you might notice a bright color, a face, or a moving object first. This is because those parts of the picture stand out to you—they are more \enquote{salient.}

A saliency map is a visual tool that shows which parts of an image are most likely to catch your attention. It highlights the areas that are most prominent or important in a picture. These maps can be created by computers using different techniques to predict where people will look when they see an image.

Here’s how it works:

\begin{itemize}
    \item \textbf{Image Analysis:} A computer program analyzes an image to find the features that stand out, like colors, edges, or unique shapes.
    \item \textbf{Highlighting Salient Areas:} The program then creates a map that highlights these features. The brighter or more colorful areas on the map show where your attention is most likely to go.
\end{itemize}

Saliency maps are not only helpful to understand what features stand out to humans but also particularly useful in understanding how computers learn to recognize images. When a computer is trained to identify objects in pictures, it looks at many examples and learns to focus on the important parts that help it make a correct identification. \textbf{The saliency map shows us which parts of the image the computer thinks are important.}

In our study, we will show you different saliency maps and measure how accurate you think the AI classification is.

\begin{enumerate}
\setcounter{enumi}{\value{qcounter}}
        \item \mychoice{I have read and understood the explanation}
\setcounter{qcounter}{\value{enumi}}
\end{enumerate}


\subsubsection{Grad-CAM}
\label{app:gradcam}
\hfill

\noindent\textbf{In this study, you will see Saliency Maps based on the Grad-CAM approach}

\noindent\emph{How to Interpret Saliency Maps Generated Using the Grad-CAM Approach}

\noindent Saliency maps generated using the Grad-CAM (Gradient-weighted Class Activation Mapping) approach are visual tools that help us understand which parts of an image a neural network (like a deep learning classifier) focuses on to make its decision. Here’s a simple explanation to interpret these maps:

\noindent What is Grad-CAM?

\noindent Grad-CAM is a technique used to explain the predictions of AI image classifiers. It produces a heatmap that shows the areas of the image that are most influential.

\noindent Reading a Saliency Map:

\begin{itemize}
    \item \textbf{Colors:} The heatmap uses colors to indicate the importance of different regions. Typically, warmer colors (reds, yellows) represent high importance, and cooler colors (blues, greens) represent low importance.
\end{itemize}

\noindent Interpreting the Map:

\begin{itemize}
    \item \textbf{High Importance Areas:}
    Look for the regions highlighted in warm colors. These are the parts of the image the model is using most to make its classification decision. For example, if the task is to classify an image of a dog, and the dog's face is highlighted in red, it indicates that the model considers the face important for making its decision.
    \item \textbf{Low Importance Areas:} Regions in cooler colors are considered less important by the model for the classification task. If the background or non-essential parts of the animal are highlighted in blue or green, it means these areas have little influence on the classification.
\end{itemize}

\noindent Assessing Classifier Performance:

\begin{itemize}
    \item \textbf{Correct Focus:} If the high importance areas correspond to the distinctive features of the animal (e.g., the face, body shape, distinctive markings or color), it suggests that the classifier is likely working well and making decisions based on relevant features.
    \item \textbf{Incorrect Focus:} If the high importance areas are unrelated to the animal, it might indicate that the classifier is not focusing on the correct parts of the image, which could lead to poor performance.
\end{itemize}

\noindent Example

\noindent Imagine you have an image of a cat, and the saliency map generated by Grad-CAM shows the following:

\begin{itemize}
    \item \textbf{High Importance (Red/Yellow):} The map highlights the cat's face and ears.
    \item \textbf{Low Importance (Blue/Green):} The map shows cooler colors around the background and the less distinctive parts of the body.
\end{itemize}

\textbf{Interpretation:} This suggests the classifier is correctly focusing on the key features that define a cat, such as the face, whiskers and ears, which are typically important for recognizing cats. Therefore, the classifier is likely working well for this image.

By looking at the saliency map, you can visually infer whether the model's attention aligns with human intuition, providing insight into how well the classifier is performing its task.

Here is an example of a Saliency Map generated using Grad-CAM (left). [See \autoref{fig:gradcam}]

\textbf{Please adjust the brightness of your monitor if you have difficulties seeing the Saliency Map.}

\begin{enumerate}
\setcounter{enumi}{\value{qcounter}}
        \item \mychoice{I have read and understood the explanation}
\setcounter{qcounter}{\value{enumi}}
\end{enumerate}


\subsubsection{Guided Backpropagation}\label{app:gbp}
\hfill

\noindent\textbf{In this study, you will see Saliency Maps based on the Guided Backpropagation approach}

\noindent\textit{How to Interpret Saliency Maps Generated Using the Guided Backpropagation Approach}

\noindent Saliency maps generated using the Guided Backpropagation approach are visual tools that help us understand which parts of an image a neural network (like a deep learning classifier) focuses on to make its decision. Here’s a simple explanation to interpret these maps:

\noindent What is Guided Backpropagation?

\noindent Guided Backpropagation is a technique used to explain the predictions of AI image classifiers. It creates a detailed map that highlights the specific features of the image that are most influential.

\noindent Reading a Saliency Map:

\begin{itemize}
    \item \textbf{Colors:} The saliency map typically uses grayscale values to indicate the importance of different regions. Brighter areas represent high importance, while darker areas represent low importance.
\end{itemize}

\noindent Interpreting the Map:

\begin{itemize}
    \item \textbf{High Importance Areas:} Look for the regions highlighted in bright colors. These are the parts of the image the model is using most to make its classification decision. For example, if the task is to classify an image of a dog, and the dog's face is highlighted in bright white, it indicates that the model considers the face important for making its decision.
    \item \textbf{Low Importance Areas:} Darker regions are considered less important by the model for the classification task. If the background or non-essential parts of the animal are darker, it means these areas have little influence on the classification.
\end{itemize}
    
\noindent Assessing Classifier Performance:

\begin{itemize}
    \item \textbf{Correct Focus:} If the high importance areas correspond to the distinctive features of the animal (e.g., the face, body shape, distinctive markings or color), it suggests that the classifier is likely working well and making decisions based on relevant features.
    \item \textbf{Incorrect Focus:} If the high importance areas are unrelated to the animal, it might indicate that the classifier is not focusing on the correct parts of the image, which could lead to poor performance.
\end{itemize}

\noindent Example

\noindent Imagine you have an image of a cat, and the saliency map generated by Guided Backpropagation shows the following:

\begin{itemize}
    \item \textbf{High Importance (Bright Areas):} The map highlights the cat's face and ears.
    \item \textbf{Low Importance (Dark Areas):} The map shows darker areas around the background and the less distinctive parts of the body.
\end{itemize}

\noindent \textbf{Interpretation:} This suggests the classifier is correctly focusing on the key features that define a cat, such as the face, whiskers and ears, which are typically important for recognizing cats. Therefore, the classifier is likely working well for this image.

By looking at the saliency map, you can visually infer whether the model's attention aligns with human intuition, providing insight into how well the classifier is performing its task.

Here is an example of a Saliency Map generated using Guided Backpropagation (left). [See \autoref{fig:gbp}]

\textbf{Please adjust the brightness of your monitor if you have difficulties seeing the Saliency Map.}

\begin{enumerate}
\setcounter{enumi}{\value{qcounter}}
        \item \mychoice{I have read and understood the explanation}
\setcounter{qcounter}{\value{enumi}}
\end{enumerate}


\subsubsection{LIME}\label{app:lime}
\hfill

\noindent\textbf{In this study, you will see Saliency Maps based on the LIME approach}

\noindent\textit{How to Interpret Saliency Maps Generated Using the LIME Approach}

\noindent Saliency maps generated using the LIME (Local Interpretable Model-agnostic Explanations) approach are visual tools that help us understand which parts of an image a neural network (or any classifier) focuses on to make its decision. Here’s a simple explanation to interpret these maps:

\noindent What is LIME?

\noindent LIME is a technique used to explain the predictions of AI image classifiers. LIME highlights which parts of the image are most influential.

\noindent Reading a Saliency Map:

\begin{itemize}
    \item \textbf{Colors:} The saliency map typically uses highlighted segments to indicate importance. Segments that are important for the classifier's decision are highlighted, while less important segments are not.
\end{itemize}

\noindent Interpreting the Map:

\begin{itemize}
    \item \textbf{High Importance Areas:} Look for the regions that are highlighted. These are the parts of the image the model is using most to make its classification decision. For example, if the task is to classify an image of a dog, and the dog's face is highlighted, it indicates that the model considers the face important for making its decision.
    \item \textbf{Low Importance Areas:} Non-highlighted regions are considered less important by the model for the classification task. If the background or non-essential parts of the animal are not highlighted, it means these areas have little influence on the classification.
\end{itemize}
    
\noindent Assessing Classifier Performance:

\begin{itemize}
    \item \textbf{Correct Focus:} If the high importance areas correspond to the distinctive features of the animal (e.g., the face, body shape, distinctive markings or color), it suggests that the classifier is likely working well and making decisions based on relevant features.
    \item \textbf{Incorrect Focus:} If the high importance areas are unrelated to the animal, it might indicate that the classifier is not focusing on the correct parts of the image, which could lead to poor performance.
\end{itemize}

\noindent Example

\noindent Imagine you have an image of a cat, and the saliency map generated by LIME shows the following:

\begin{itemize}
    \item \textbf{High Importance (Bright Areas):} The map highlights the cat's face and ears.
    \item \textbf{Low Importance (Dark Areas):} The map shows darker areas around the background and the less distinctive parts of the body.
\end{itemize}

\noindent \textbf{Interpretation:} This suggests the classifier is correctly focusing on the key features that define a cat, such as the face, whiskers and ears, which are typically important for recognizing cats. Therefore, the classifier is likely working well for this image.

By looking at the saliency map, you can visually infer whether the model's attention aligns with human intuition, providing insight into how well the classifier is performing its task.

Here is an example of a Saliency Map generated using LIME (left). [See \autoref{fig:gbp}]

\textbf{Please adjust the brightness of your monitor if you have difficulties seeing the Saliency Map.}

\begin{enumerate}
\setcounter{enumi}{\value{qcounter}}
        \item \mychoice{I have read and understood the explanation}
\setcounter{qcounter}{\value{enumi}}
\end{enumerate}

\subsubsection{Comprehension Questions}\label{app:comprehension}\hfill

\noindent\textbf{Please answer the following questions before starting the experiment}

It is very important that you understood the explanations regarding saliency maps. Please answer the following questions.

You can use the \enquote{back} button at the end of the page to re-read the explanations the questions are based on.

If you answer incorrectly, you will jump back in the questionnaire and see the explanations again. If you answer incorrectly twice in a row, you will be asked to return your submission on Prolific and the questionnaire will end.

\begin{enumerate}
\setcounter{enumi}{\value{qcounter}}
        \item \textbf{Based on the explanation provided, which of the following best describes the purpose of a saliency map in the context of this study? (Only one is correct)}
            \mychoice{A saliency map is a tool used to edit images by enhancing their colors and shapes.}
            \mychoice{A saliency map is a visual representation that shows which parts of an image are most important for an AI-based classification.}
            \mychoice{A saliency map highlights the least important parts of an image.}
            \mychoice{A saliency map is a method to measure how quickly an AI can process an image.}
        \item \textbf{Which of the following statements best explains how to interpret high importance areas on a saliency map generated using Guided Backpropagation? (Only one is correct)}
            \mychoice{High importance areas are the darkest regions of the map, indicating the features least used by the model for classification.}
            \mychoice{High importance areas are the brightest regions of the map, indicating the features most used by the model for classification.}
            \mychoice{High importance areas are highlighted in color to show the most aesthetically pleasing parts of the image.}
            \mychoice{High importance areas are blurred to indicate where the model is uncertain about the classification.}
        \item \textbf{What does it suggest about a classifier’s performance if the saliency map highlights the background of an image as highly important? (Only one is correct)}
            \mychoice{The classifier is performing well and correctly focusing on the relevant features of the image.}
            \mychoice{The classifier is not performing well and is focusing on irrelevant features of the image.}
            \mychoice{The classifier is indicating areas that need more training data.}
            \mychoice{The classifier is highlighting the most aesthetically pleasing parts of the image.}
\setcounter{qcounter}{\value{enumi}}
\end{enumerate}

\subsubsection{User task}
\label{app:usertask}

\begin{figure*}[!htbp]
    \centering
    \includegraphics[width=0.75\linewidth]{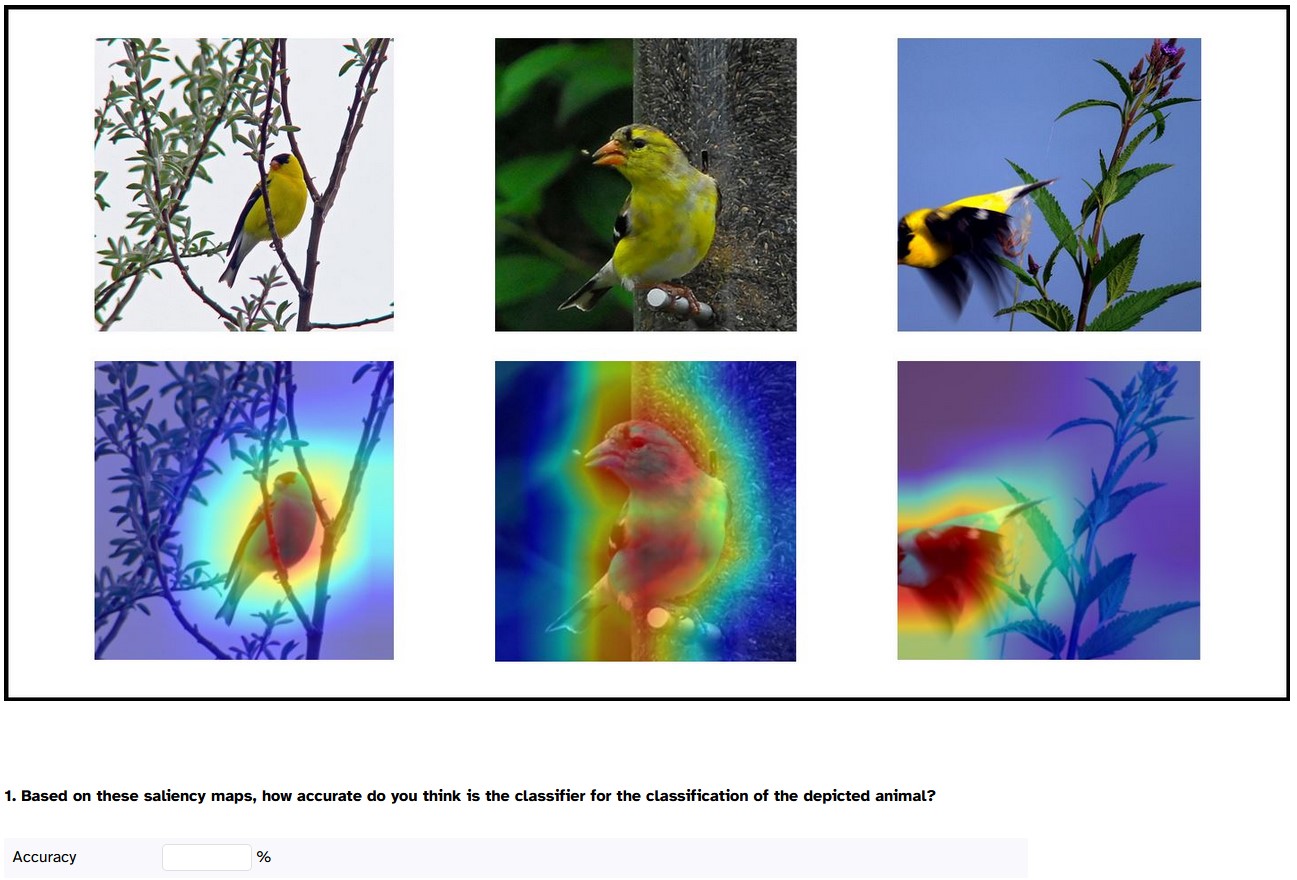}  
    \caption{Example of the Grad-CAM task}
    \Description{Example of the Grad-CAM task}
    \label{Grad-CAM task}
\end{figure*}

\begin{figure}[!htbp]
    \centering
    \includegraphics[width=0.75\linewidth]{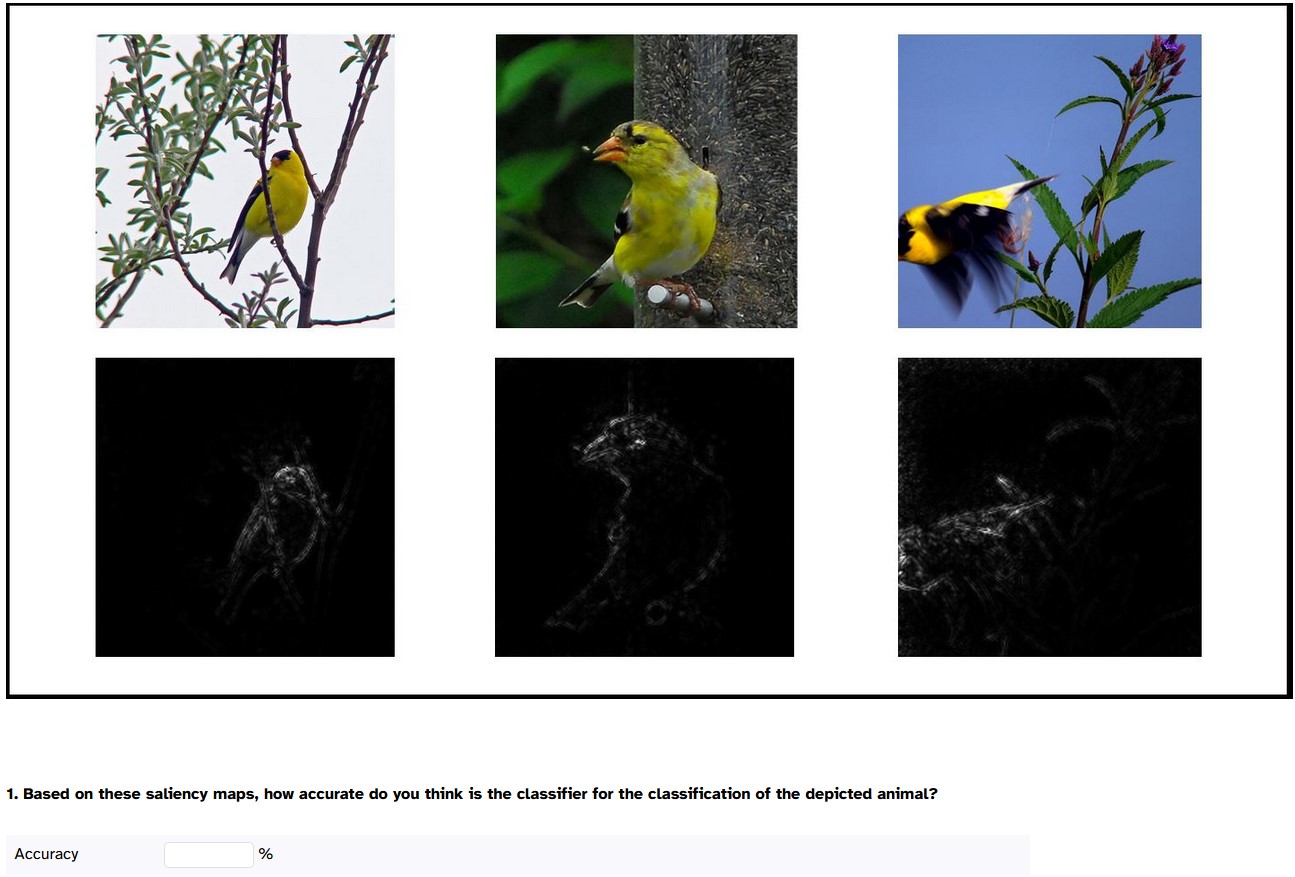}  
    \caption{Example of the GBP task}
    \Description{Example of the GBP task}
    \label{GBP task}
\end{figure}

\begin{figure}[!htbp]
    \centering
    \includegraphics[width=0.75\linewidth]{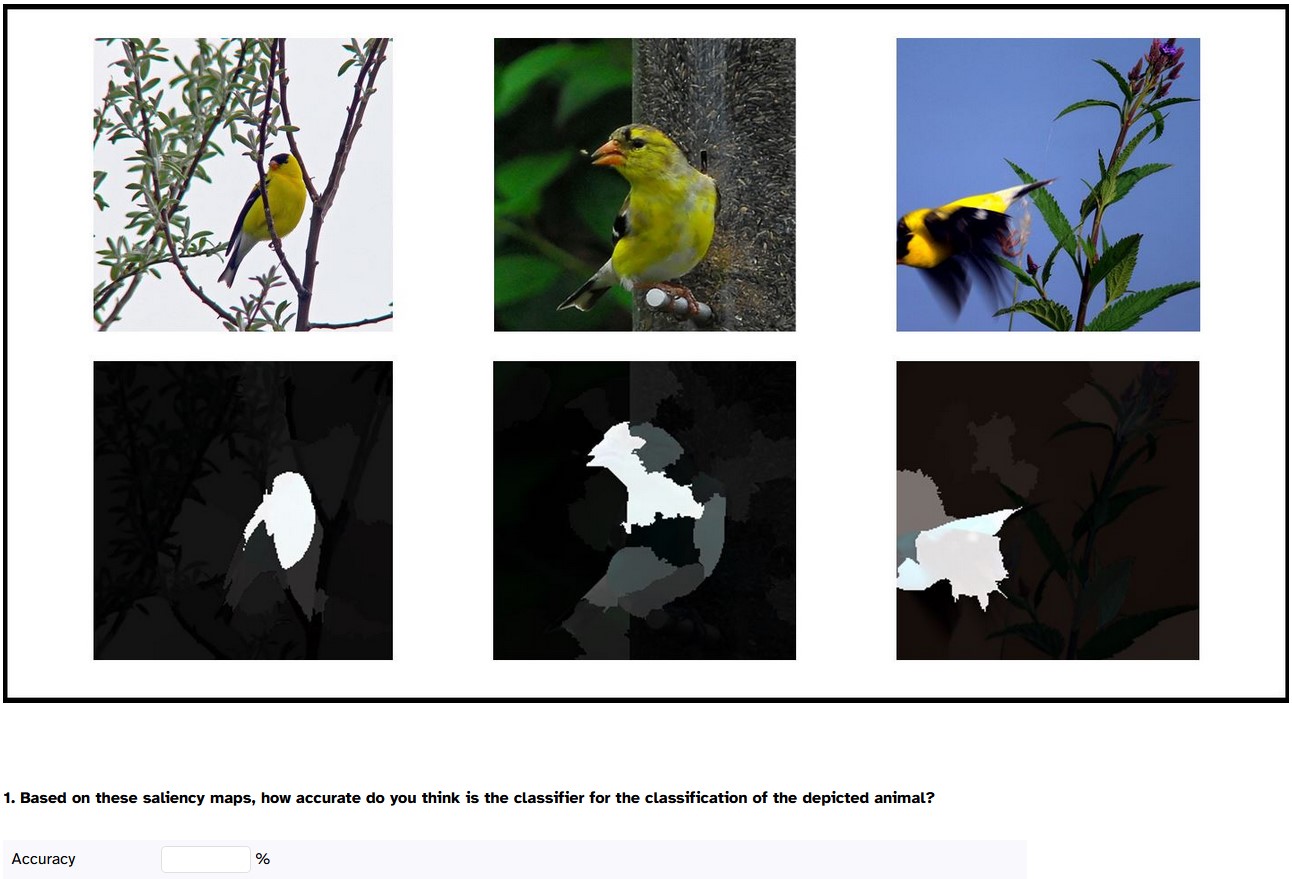}  
    \caption{Example of the LIME task}
    \Description{Example of the LIME task}
    \label{LIME task}
\end{figure}

\begin{figure}[!htbp]
    \centering
    \includegraphics[width=0.75\linewidth]{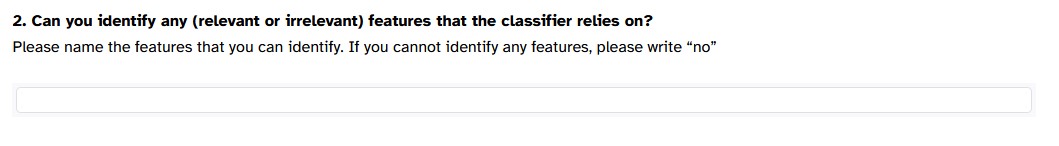}  
    \caption{Continuation of the user task (identical in all groups)}
    \Description{Continuation of the user task (identical in all groups)}
    \label{task 2}
\end{figure}

\begin{figure}[!htbp]
    \centering
    \includegraphics[width=0.75\linewidth]{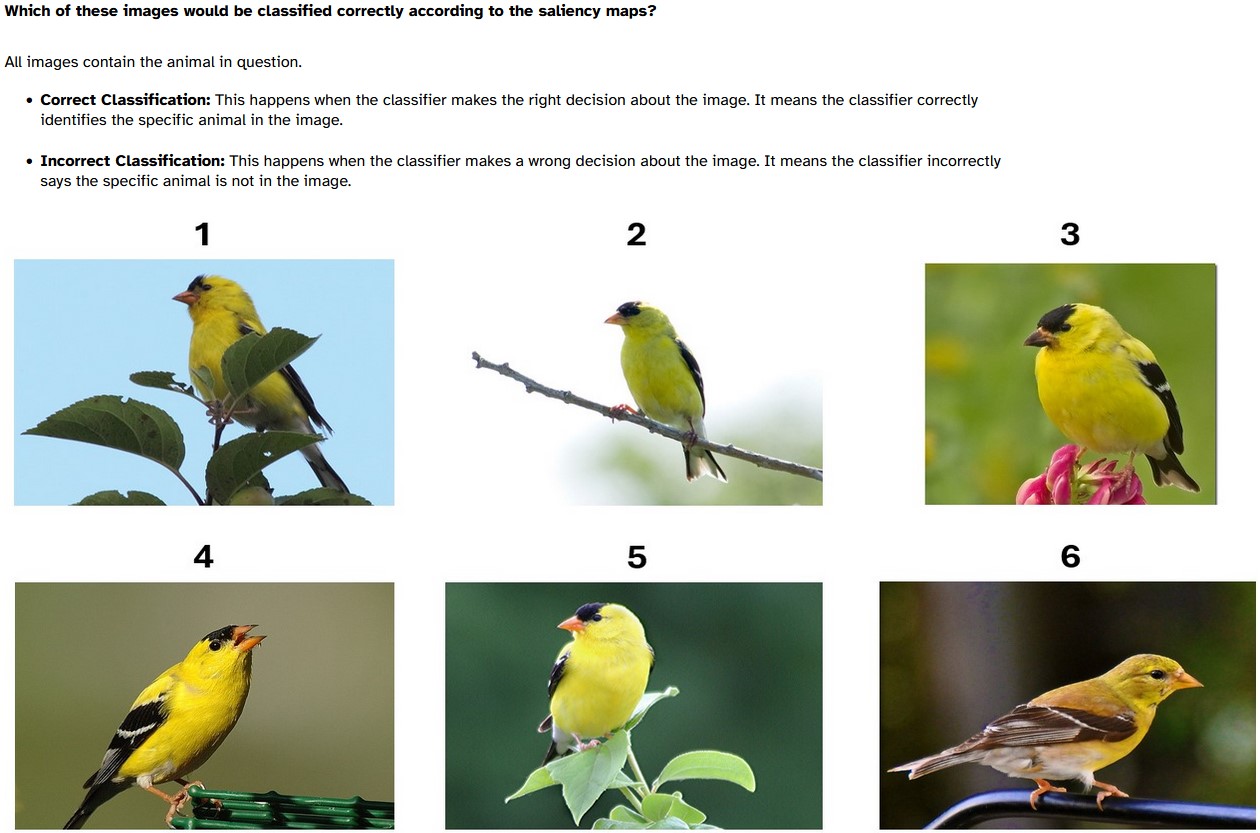}  
    \caption{Continuation of the user task (identical in all groups)}
    \Description{Continuation of the user task (identical in all groups)}
    \label{task 3}
\end{figure}

\begin{figure}[!htbp]
    \centering
    \includegraphics[width=0.75\linewidth]{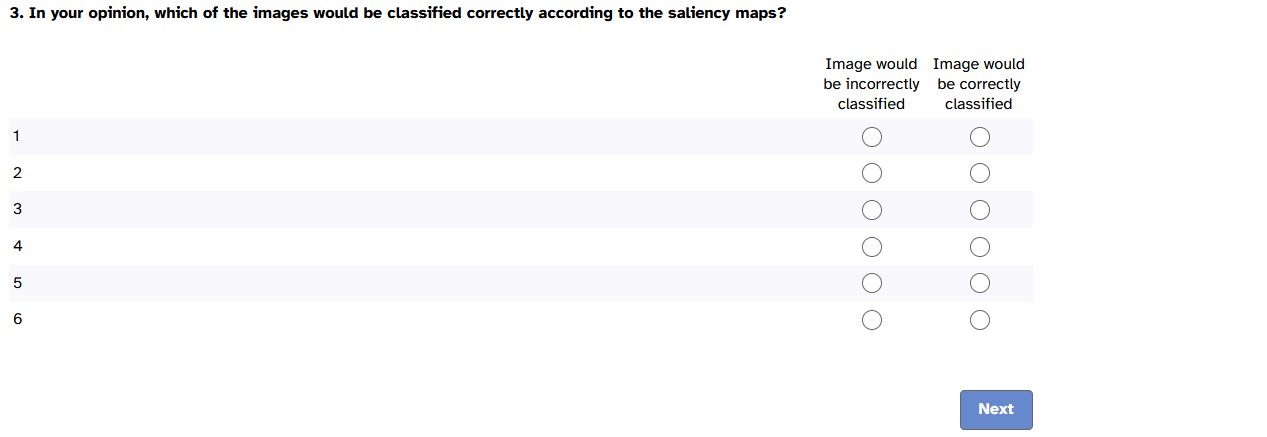}  
    \caption{Continuation of the user task (identical in all groups)}
    \Description{Continuation of the user task (identical in all groups)}
    \label{task 4}
\end{figure}

\section{Standard Deviations of Mathematical Metrics}

\begin{table*}[h!]
\centering
\setlength{\tabcolsep}{2pt}
\begin{tabular}{l ccc cc c c}
\toprule
\textbf{Group} & \multicolumn{3}{c}{Fidelity} & \multicolumn{2}{c}{Complexity} & Robustness & Localization \\
\cmidrule(lr){2-4} \cmidrule(lr){5-6} \cmidrule(lr){7-7} \cmidrule(lr){8-8}
& {Insertion} & {Deletion} & {Monotonicity} & {Sparseness} & {Ent. Complexity} & {Avg. Sensitivity} & {EPG} \\
\midrule
\textbf{Grad-CAM} 
& 0.0977 
& 0.0672 
& 0.1644 
& 0.0567 
& 0.0814 
& 0.0095 
& 0.1309 \\

\textbf{GBP}     
& 0.1344 
& 0.0475 
& 0.1019 
& 0.0429 
& 0.2309 
& 0.4101 
& 0.1687 \\

\textbf{LIME}    
& 0.1234 
& 0.0564 
& 0.2149 
& 0.0716 
& 0.2142 
& 0.0143 
& 0.1589 \\
\bottomrule
\end{tabular}
\caption{Standard deviations of the mathematical properties across 12 trials for each explanation method.}
\label{tab:sd_properties}
\end{table*}

\begin{table}[htbp]
\setlength{\tabcolsep}{10pt}
\centering
\caption{Results of the linear mixed models examining the association between mathematical saliency map properties and participants' absolute accuracy differences (RQ4). All coefficients are standardized regression coefficients ($\beta$), obtained after standardizing both the mathematical metric predictors and the dependent variable. Positive Beta values indicate that higher values of the respective metric are associated with larger deviations between participants' estimated and the true classifier accuracy, whereas negative values indicate smaller deviations. Confidence intervals (CI) represent $95\%$ Wald confidence intervals. Each model included one mathematical metric as a fixed effect and a random intercept for participants. The analyses included data from all $N=166$ participants.}
\label{tab:s1}
\begin{tabular}{lllccccc}
\toprule
\textbf{Group}& Metric & Saliency & $ \beta$ & CI & SE & $t$ & $p$ \\
\midrule
\mr{9}{Fidelity}&\mr{3}{Insertion $\uparrow$} & \textbf{GBP}          & 0.005  & [-0.070, 0.079]  & 0.038 & 0.12  & 0.903 \\
&&\textbf{LIME}          & 0.082  & [0.019, 0.146]   & 0.033 & 2.53  & 0.012 \\
&&\textbf{GradCAM}       & -0.059 & [-0.154, 0.035]  & 0.048 & -1.23 & 0.220 \\
\cmidrule{2-8}
&\mr{3}{Deletion $\downarrow$}&\textbf{GBP}            & 0.196  & [0.088, 0.305]   & 0.055 & 3.55  & $< .001$ \\
&&\textbf{LIME}           & 0.110  & [0.037, 0.183]   & 0.037 & 2.95  & 0.003 \\
&&\textbf{GradCAM}        & 0.011  & [-0.061, 0.083]  & 0.037 & 0.29  & 0.772 \\
\cmidrule{2-8}
&\mr{3}{Monotonicity $\uparrow$}& \textbf{GBP}     & 0.538  & [0.355, 0.720]   & 0.093 & 5.78  & $< .001$ \\
&& \textbf{LIME}    & 0.014  & [-0.056, 0.085]  & 0.036 & 0.40  & 0.691 \\
&& \textbf{GradCAM} & -0.092 & [-0.200, 0.016]  & 0.055 & -1.67 & 0.096 \\
\midrule
\mr{6}{Complexity}&\mr{3}{{Sparseness $\uparrow$}}& \textbf{GBP}     & -0.547 & [-0.861, -0.233] & 0.160 & -3.41 & $< .001$ \\
&& \textbf{LIME}    & -0.157 & [-0.308, -0.007] & 0.077 & -2.05 & 0.041 \\
&& \textbf{GradCAM} & 0.165  & [-0.057, 0.387]  & 0.113 & 1.45  & 0.146 \\
\cmidrule{2-8}
&\mr{3}{Ent.Complexity $\downarrow$}& \textbf{GBP}     & 0.338  & [0.153, 0.523]   & 0.094 & 3.59  & $< .001$ \\
&& \textbf{LIME}    & 0.052  & [-0.109, 0.212]  & 0.082 & 0.63  & 0.528 \\
&& \textbf{GradCAM} & -0.571 & [-1.061, -0.082] & 0.250 & -2.29 & 0.023 \\
\midrule
\mr{3}{Robustness}&\mr{3}{Avg. Sensitivity $\downarrow$}& \textbf{GBP}     & -0.050 & [-0.136, 0.036]  & 0.044 & -1.14 & 0.255 \\
&& \textbf{LIME}    & -2.908 & [-4.850, -0.966] & 0.991 & -2.93 & 0.003 \\
&& \textbf{GradCAM} & -9.350 & [-12.717, -5.983]& 1.718 & -5.44 & $< .001$ \\
\midrule
\mr{3}{Localization}&\mr{3}{EPG $\uparrow$}& \textbf{GBP}     & -0.196 & [-0.270, -0.123] & 0.038 & -5.23 & $< .001$ \\
&& \textbf{LIME}    & -0.059 & [-0.123, 0.004]  & 0.032 & -1.83 & 0.067 \\
&& \textbf{GradCAM} & -0.152 & [-0.242, -0.063] & 0.046 & -3.34 & $< .001$ \\
\bottomrule
\end{tabular}
\end{table}

\end{document}